\documentclass[conference]{IEEEtran}
\IEEEoverridecommandlockouts
\usepackage{cite}
\usepackage{amsmath,amssymb,amsfonts}
\usepackage{graphicx}
\usepackage{textcomp}
\usepackage{xcolor}
\usepackage{algorithm}
\usepackage[noend]{algorithmic}
\usepackage{bm}
\usepackage{url}
\usepackage{multirow}
\usepackage{tabularx}
\usepackage{enumerate}

\newcommand{\figscalea}{0.19}

\newcommand{\figheightvisori}{3.5cm}
\newcommand{\figheightvis}{3.5cm}
\newtheorem{lemma}{Lemma}

\newcolumntype{A}{>{\raggedright\arraybackslash}p{1.25cm}}
\newcolumntype{B}{>{\centering\arraybackslash}p{1.5cm}}
\newcolumntype{C}{>{\centering\arraybackslash}p{0.75cm}}
\newcolumntype{D}{>{\raggedright\arraybackslash}p{1.25cm}}
\newcolumntype{E}{>{\centering\arraybackslash}p{1.9cm}}
\newcolumntype{F}{>{\centering\arraybackslash}p{0.6cm}}
\newcolumntype{G}{>{\centering\arraybackslash}p{1.8cm}}
\newcolumntype{H}{>{\centering\arraybackslash}p{1.3cm}}
\newcolumntype{I}{>{\centering\arraybackslash}p{1cm}}

\def\BibTeX{{\rm B\kern-.05em{\sc i\kern-.025em b}\kern-.08em
    T\kern-.1667em\lower.7ex\hbox{E}\kern-.125emX}}
\begin{document}

\title{Social Graph Restoration via Random Walk Sampling}

\author{
\IEEEauthorblockN{Kazuki Nakajima}
\IEEEauthorblockA{\textit{Tokyo Institute of Technology} \\
nakajima.k.an@m.titech.ac.jp}
\and
\IEEEauthorblockN{Kazuyuki Shudo}
\IEEEauthorblockA{\textit{Tokyo Institute of Technology} \\
shudo@is.titech.ac.jp}
}

\maketitle

\begin{abstract}
Analyzing social graphs with limited data access is challenging for third-party researchers.
To address this challenge, a number of algorithms that estimate structural properties via a random walk have been developed.
However, most existing algorithms are limited to the estimation of local structural properties. 
Here we propose a method for restoring the original social graph from the small sample obtained by a random walk.
The proposed method generates a graph that preserves the estimates of local structural properties and the structure of the subgraph sampled by a random walk.
We compare the proposed method with subgraph sampling using a crawling method and the existing method for generating a graph that structurally resembles the original graph via a random walk.
Our experimental results show that the proposed method more accurately reproduces the local and global structural properties on average and the visual representation of the original graph than the compared methods.
We expect that our method will lead to exhaustive analyses of social graphs with limited data access.
\end{abstract}


\section{Introduction}
Studies on online social networks have played an important role in understanding social characteristics, such as human connections and behaviors, on a worldwide scale.
A number of studies, e.g., \cite{ahn, mislove, kwak, backstrom2012, traud2012}, have investigated the structural properties of social graphs, wherein nodes represent users and edges represent friendships between users in online social networks.
In general, researchers sample the graph data for analysis because all the graph data are not available to third-party researchers.
Crawling methods in which one repeatedly traverses a neighbor (e.g., breadth-first search and random walk) are effective for sampling graph data in online social networks where the neighbors' data of a user is available by querying the user \cite{ahn, mislove, kwak, wilson, gjoka_walking, catanese, gjoka_practical}. 

Gjoka et al. proposed a framework called re-weighted random walk to obtain unbiased estimates of the structural properties of social graphs via a random walk \cite{gjoka_practical, gjoka_walking}.
This framework addresses the sampling bias toward high-degree nodes typically induced by crawling methods. 
First, one performs a random walk on the graph (i.e., one repeatedly moves to a neighbor chosen uniformly and randomly) to obtain a sequence of sampled nodes.
Then, one re-weights each sampled node to correct the sampling bias derived from the Markov property of the sequence.
The number of available queries within a practical sampling period is typically limited \cite{katzir_nodecc, dasgupta, chiericetti}.
Therefore, a number of algorithms that estimate structural properties using a small number of queries based on this framework have been developed \cite{ribeiro, katzir_node, gjoka_2_5_k, katzir_nodecc, dasgupta, wang, chen, han, nakajima_jip, bera}.

However, re-weighted random walk enables analysts only to estimate local structural properties in principle.
First, this framework forces analysts to sample most graph data to correct the sampling bias when attempting to estimate global structural properties, such as the shortest-path properties. 
Second, the quantity of re-weighted sample means is not sufficient to predict the structure of the original graph, such as its visual representation.
On the other hand, analysts' interests in the characteristics of social networks are generally diverse \cite{borgatti}; these characteristics include local structural properties (e.g., the degree distribution and clustering coefficient), global structural properties (e.g., the distributions of shortest-path lengths and betweenness centrality), and visual graph representations.

To address this gap, we study the social graph restoration problem: Given a small sample of a social graph obtained by crawling, we aim to generate a graph whose structural properties are as close as possible to the corresponding properties of the original graph.
The generated graph enables us to estimate local and global structural properties and predict the visual representation of the original graph.
Existing methods to address this problem include subgraph sampling \cite{lee, leskovec, ahn, mislove, wilson, kwak, catanese, ahmed} and Gjoka et al.'s method  \cite{gjoka_2_5_k}.
In subgraph sampling, one constructs the subgraph induced from a set of edges obtained using a crawling method and implicitly assumes that the subgraph is a representative sample of the original graph.
In contrast, Gjoka et al.'s method generates a graph that preserves estimates of local structural properties obtained by re-weighted random walk, intending to reproduce the structural properties of the original graph, including those that are not intended to be preserved.

In this paper, we propose a method for restoring the original social graph from the small sample obtained by a random walk.
The proposed method generates a graph that preserves the estimates of local structural properties and the structure of the subgraph sampled by a random walk. 
First, we construct the subgraph induced from a set of edges obtained during a random walk and estimate the local structural properties using re-weighted random walk.
Then, we complement the nodes and edges in the subgraph to generate a graph that preserves the estimates of local properties.
To realize this, we extend a family of graph-generative models called $dK$-series \cite{mahadevan, gjoka_2_5_k, orsini} which attempts to reproduce various structural properties of a given graph by preserving the local structural properties.
We conduct experiments using seven real social graph datasets to evaluate the effectiveness of the proposed method.
The experimental results show that the proposed method outperforms the existing methods in terms of the average accuracy of 12 structural properties and the visual representation of the generated graphs.
We also find that the generation time of the proposed method is several times faster than that of Gjoka et al.'s method.
The source code for our method is available at \url{https://github.com/kazuibasou/social-graph-restoration}.

\section{Related Work}
Gjoka et al. designed a framework called re-weighted random walk to obtain the unbiased estimates of structural properties via a random walk \cite{gjoka_practical,gjoka_walking}.
In the past decade, a number of algorithms based on this framework have been developed for accurately estimating structural properties using a small number of queries.
Examples of the structural properties include the number of nodes \cite{katzir_node, katzir_nodecc}, average degree \cite{dasgupta, gjoka_walking, gjoka_practical}, degree distribution \cite{gjoka_walking, gjoka_practical}, joint degree distribution \cite{gjoka_2_5_k}, clustering coefficients \cite{katzir_nodecc, ribeiro, bera}, motifs and graphlets \cite{chen, wang, han}, and node centrality \cite{nakajima_jip}.
However, most of the existing algorithms are limited to the estimation of local structural properties.
In this work, we study another problem called social graph restoration in which we aim to generate a graph that structurally resembles the original graph from the small sample obtained using a crawling method.

We regard subgraph sampling as a baseline method for the social graph restoration problem.
In subgraph sampling, one constructs the subgraph induced from a set of edges obtained using a crawling method \cite{lee, leskovec, ahmed}.
In early studies, the subgraph was implicitly assumed to be a representative sample of the original graph \cite{mislove, ahn, wilson, kwak, catanese}. 
However, Gjoka et al. demonstrated that crawling methods typically introduce a significant sampling bias toward high-degree nodes \cite{gjoka_practical,gjoka_walking}.
In this study, we compare the proposed method with subgraph sampling using each of the well-used crawling methods (i.e., breadth-first search \cite{mislove, wilson, kurant2011, catanese,  rozemberczki2020}, snowball sampling \cite{goodman1961, lee2006, ahn, illenberger2012, rozemberczki2020}, forest fire sampling \cite{leskovec, de2010, ahmed, rozemberczki2020}, and random walk).
We confirm that subgraph sampling using a small sample typically introduces bias in structural properties on average and misses the surrounding structure that consists of low-degree nodes in the graph visualization.

Gjoka et al. proposed a method for generating a graph that preserves the estimates of the joint degree distribution and degree-dependent clustering coefficient obtained by re-weighted random walk \cite{gjoka_2_5_k}.
The authors showed that the generated graph accurately reproduces not only local structural properties but also global structural properties that are not intended to be preserved.
In this study, we propose a method for generating a graph that preserves both the estimates of local structural properties (i.e., the number of nodes, average degree, degree distribution, joint degree distribution, and degree-dependent clustering coefficient) and the structure of the subgraph sampled by a random walk. 
Specifically, we add nodes and edges to the subgraph sampled by a random walk to ensure that the final graph preserves the estimates of local structural properties. 
Our underlying idea is to optimally use the raw structural information of the subgraph in the generation process.
Our experimental results show that the proposed method more accurately reproduces an average of 12 structural properties and the visual representation of the original graph and has a generation time that is several times faster than that of Gjoka et al.'s method.

Several studies have developed random walk algorithms to improve the accuracy of estimators or the efficiency of the number of queries \cite{ribeiro, lee, li, nazi, zhou, li_2019, nakajima_kdd, yi2021}.
Ribeiro and Towsley proposed multidimensional random walks, which improve the estimation accuracy over a simple random walk (i.e., one repeatedly moves to a neighbor chosen uniformly and randomly on the graph) in the presence of disconnected connected components \cite{ribeiro}.
Lee et al. proposed the non-backtracking random walk algorithm, which improves the query efficiency while preserving the Markov property of the sample sequence \cite{lee}.
Nakajima and Shudo recently proposed a random walk algorithm to reduce the bias caused by private nodes whose neighbors' data are not retrievable in social networks \cite{nakajima_kdd}.
In this study, we propose a method for restoring the original social graph via a simple random walk. 
However, while it is not trivial, it is possible to combine the above improved random walks with the proposed method.

Graph-generative models have been developed to reproduce the structural properties of a given graph \cite{chakrabarti, mahadevan, robins, leskovec_kronecker, goldenberg, orsini, netgan, you, vankoevering2021}. 
In this study, we extend a family of generative models called the $dK$-series \cite{mahadevan, gjoka_2_5_k, orsini} to the generation of a graph that preserves the estimates of local structural properties and the structure of the subgraph sampled by a random walk. 
It is not trivial to extend any generative model including the $dK$-series, which assumes that all graph data are available, to the social graph restoration problem because of the following three reasons.
First, we need to estimate the input parameters of the model from a sample of the original graph.
Second, we need to construct the input parameters from their estimates so that those parameters meet all conditions required to realize the desired graph.
Third, although one adds nodes and edges in an empty graph in most generative models, we add nodes and edges to the sampled subgraph to restore the original graph.

Social graph restoration is related to matrix completion \cite{candes}, in which one complements entries in a given matrix, and is also related to link prediction \cite{liben},  network completion \cite{kim}, and network inference \cite{huisman2009, handcock2010, newman2018}, in which one complements nodes or edges in a given graph.
However, the subgraph sampled by a random walk omits nodes biased toward low degrees and their edges, which is a special case of the assumption of these problems regarding missing nodes or edges.
The proposed method is specialized in the case of complementing the nodes and edges in the subgraph sampled by a random walk.

\section{Preliminaries}
\subsection{Problem definition}
We represent a connected and undirected social graph as $\mathcal{G} = (\mathcal{V}, \mathcal{E})$, where $\mathcal{V} = \{v_1, v_2, \ldots, v_n\}$ is a set of nodes (users) and $\mathcal{E}$ is a set of edges (friendships). 
We allow multiple edges and loops. 
Let $n$ denote the number of nodes and $m$ denote the number of edges.
We denote the adjacency matrix for $\mathcal{G}$ by $\bm{A}$.
We assume that $A_{ij}$ is the number of edges between $v_i$ and $v_j$ $(i \neq j)$.
We assume that $A_{ii}$ is equal to twice the number of loops of $v_i$ by convention \cite{newman_networks}.
Let $\mathcal{N}(i)$ denote a set of edges connected to $v_i$.
Let $d_i = |\mathcal{N}(i)|$ be the degree of $v_i$ and $k_{\text{max}}$ be the maximum degree of the node.
Let $1_{\{cond\}}$ denote a function that returns 1 if a condition $cond$ holds and 0 otherwise.

We assume the standard model for accessing graph $\mathcal{G}$ as in Refs. \cite{ahn, mislove, wilson, gjoka_walking, kwak, gjoka_practical}: 
(i) if one queries node $v_i$, the set $\mathcal{N}(i)$ is available; 
(ii) completely or randomly accessing $\mathcal{G}$ is not feasible; and
(iii) the graph $\mathcal{G}$ is static.
Crawling methods (e.g., breadth-first search and random walk) are effective for sampling the nodes and edges of a graph in this access model.

We study the following problem: given a sampling list of the indices and connected edges of a small fraction of nodes queried using a crawling method, we generate a graph whose structural properties are as close as possible to the corresponding structural properties of the original graph $\mathcal{G}$.

\subsection{Random walk}

We obtain a sequence of sampled nodes via a simple random walk as follows.
We select a seed node $v_{x_1}$, where $x_i$ denotes the index of the $i$-th sampled node.
For the $i$-th sampled node ($i = 1, \ldots, r-1$), we select an edge uniformly at random from the set $\mathcal{N}(x_i)$ and then pass through the edge.
Finally, we obtain a list of the indices and connected edges of $r$ sampled nodes, as denoted by $\mathcal{L} = ((x_i, \mathcal{N}(x_i)))_{i=1}^r$.

\subsection{$dK$-series}
We use the concept of a family of graph-generative models called the $dK$-series \cite{mahadevan, gjoka_2_5_k, orsini}. 
The $dK$-series defines a series of random graphs called $dK$-graphs that preserve all the joint degree distributions of the nodes in the subgraphs of size $d$ or less in a given graph.
$0K$-graphs preserve the number of nodes $n$ and the average degree $\bar{k}$ of a given graph, where we define
\begin{align}
\bar{k} = \frac{2m}{n}.
\label{eq:1}
\end{align}
$1K$-graphs preserve $n$, $\bar{k}$, and the degree distribution $\{P(k)\}_k$ of a given graph.
We define
\begin{align}
P(k) = \frac{n(k)}{n}
\label{eq:2}
\end{align}
for $k = 1, \ldots, k_{\text{max}}$, where $n(k)$ is the number of nodes with degree $k$.
Preserving $n$, $\bar{k}$, and $\{P(k)\}_k$ is identical to preserving $\{n(k)\}_k$.
We refer to $\{n(k)\}_k$ as a {\it degree vector}, as in Ref. \cite{stanton}.
$2K$-graphs preserve $n$, $\bar{k}$, $\{P(k)\}_k$, and the joint degree distribution $\{P(k, k')\}_{k, k'}$ of a given graph.
We define
\begin{align}
P(k, k') = \frac{\mu(k, k') m(k, k')}{2m}
\label{eq:3}
\end{align}
for $k = 1, \ldots, k_{\text{max}},\ k' = 1, \ldots, k_{\text{max}}$, where $m(k, k')$ is the number of edges between nodes with degree $k$ and nodes with degree $k'$.
We define $\mu(k, k') = 1$ if $k \neq k'$ and $\mu(k, k) = 2$ otherwise such that $P(k, k')$ is normalized; i.e., $\sum_{k=1}^{k_{\text{max}}} \sum_{k'=1}^{k_{\text{max}}} P(k, k') = 1$.
Preserving $n$, $\bar{k}$, $\{P(k)\}_k$, and $\{P(k, k')\}_{k, k'}$ is identical to preserving $\{m(k, k')\}_{k, k'}$. 
We refer to $\{m(k, k')\}_{k, k'}$ as a {\it joint degree matrix}, as in Ref. \cite{stanton}.
$2.5K$-graphs preserve $n$, $\bar{k}$, $\{P(k)\}_k$, $\{P(k, k')\}_{k, k'}$, and the degree-dependent clustering coefficient $\{\bar{c}(k)\}_k$ of a given graph.
For $k = 1, \ldots, k_{\text{max}}$, we define
\begin{align*}
\bar{c}(k) = \frac{1}{n(k)} \sum_{i=1,\ d_i = k}^n \frac{2 t_i}{k(k-1)},
\end{align*}
where $t_i = \sum_{j=1,\ j \neq i}^{n-1} \sum_{l=j+1,\ l \neq i}^n A_{ij} A_{il} A_{jl}$ is the number of triangles to which $v_i$ belongs and $\bar{c}(1) = 0$.

$dK$-graphs more accurately reproduce the structural properties of a given graph as the value of $d$ increases \cite{mahadevan, orsini}.
Gjoka et al. demonstrated that $2.5K$-graphs successfully reproduce not only local structural properties but also global structural properties (e.g., shortest path properties) that are not intended to be preserved \cite{gjoka_2_5_k}.

\begin{figure}[t]
        \begin{center}
          \includegraphics[scale=0.36]{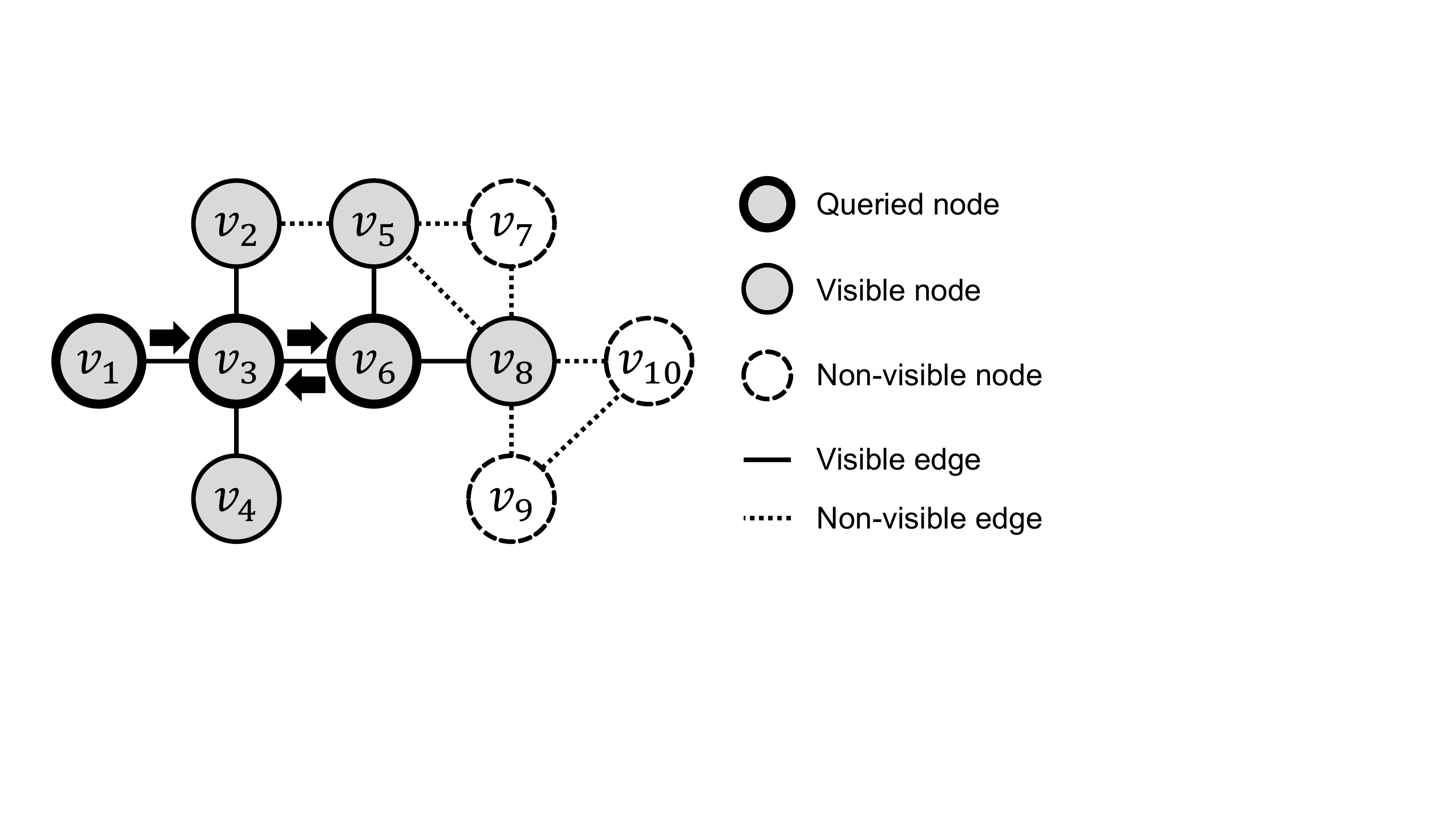}
        \end{center}
      \caption{An example of a random walk on a graph.}
      \label{fig:1}
\end{figure}

\subsection{Subgraph sampling} \label{section:3.C}
In our method, we first construct the subgraph induced from a set of edges obtained using a random walk.
We find a subset of edges in $\mathcal{G}$ obtained through a random walk as
\begin{align*}
\mathcal{E}' = \bigcup_{v_i \in \mathcal{V}'_{\text{qry}}} \mathcal{N}(i),
\end{align*}
where $\mathcal{V}'_{\text{qry}}$ denotes a set of queried nodes.
In other words, $\mathcal{E}'$ is a union set of edges connected to each of queried nodes.
Then, we construct the subgraph, $\mathcal{G}' = (\mathcal{V}', \mathcal{E}')$, which is induced from the subset $\mathcal{E}'$. 
The subset $\mathcal{V}'$ consists of two disjoint sets, i.e., $\mathcal{V}'_{\text{qry}}$ and $\mathcal{V}'_{\text{vis}}$, where $\mathcal{V}'_{\text{vis}}$ denotes a set of nodes visible as neighbors of the queried nodes. 

Figure \ref{fig:1} shows an example in which we traverse nodes in the order $v_1$, $v_3$, $v_6$, and $v_3$ via a random walk on a graph.
In this example, we obtain $\mathcal{G}' = (\mathcal{V}', \mathcal{E}')$, where $\mathcal{V}'_{\text{qry}} = \{v_1, v_3, v_6\}$, $\mathcal{V}'_{\text{vis}} = \{v_2, v_4, v_5, v_8\}$, $\mathcal{V}' = \{v_1, v_2, v_3, v_4, v_5, v_6, v_8\}$ and $\mathcal{E}' = \{(v_1, v_3), (v_2, v_3), (v_3, v_4), (v_3, v_6), (v_5, v_6), (v_6, v_8)\}$.

\subsection{Unbiased estimators of local structural properties} \label{section:3.C}
Then, we estimate the number of nodes, average degree, degree distribution, joint degree distribution, and degree-dependent clustering coefficient of the original graph from the sampling list $\mathcal{L}$.
For this purpose, we use existing estimators based on re-weighted random walk as follows.

Let $\mathcal{I} = \{(i, j)\ |\ M \leq |i - j| \land 1 \leq i,\ j \leq r\}$ denote a set of integer pairs that are between $1$ and $r$ and are located at least a threshold $M$ away. 
An unbiased estimator of the number of nodes \cite{katzir_node, katzir_nodecc} is given by 
\begin{align*}
\hat{n} \triangleq \frac{\sum_{(i,j) \in \mathcal{I}} d_{x_{i}}/d_{x_{j}}}{\sum_{(i,j) \in \mathcal{I}} 1_{\{x_{i}=x_{j}\}}}.
\end{align*}
We set $M = 0.025 r$, as in the previous study \cite{katzir_nodecc}.

An unbiased estimator of the average degree \cite{gjoka_practical, dasgupta} is given by $\hat{\bar{k}} \triangleq 1/\bar{\Phi}$, where we define 
\begin{align*}
\bar{\Phi} = \frac{1}{r} \sum_{i=1}^r 1/d_{x_i}.
\end{align*}
An unbiased estimator of the degree distribution \cite{ribeiro, gjoka_walking, gjoka_practical} is given by $\hat{P}(k) \triangleq \Phi(k)/\bar{\Phi}$, where we define 
\begin{align*}
\Phi(k) = \frac{1}{kr} \sum_{i=1}^r 1_{\{d_{x_i} = k\}}.
\end{align*}

An unbiased estimator of the joint degree distribution is given by combining the following two methods \cite{gjoka_2_5_k}: induced edges (IE) and traversed edges (TE).
The unbiased estimator of the joint degree distribution using IE is defined as $\hat{P}_{\text{IE}}(k, k') \triangleq \hat{n} \hat{\bar{k}} \Phi(k, k')$, where we define 
\begin{align*}
\Phi(k, k') = \frac{1}{k k'|\mathcal{I}|} \sum_{(i,j) \in \mathcal{I}} 1_{\{d_{x_i}=k \land d_{x_j}=k'\}} A_{x_i x_j}.
\end{align*}
Then, the unbiased estimator of the joint degree distribution using TE is defined as 
\begin{align*}
&\hat{P}_{\text{TE}}(k, k') \\
\triangleq &\frac{1}{2(r-1)} \sum_{i=1}^{r-1} &(1_{\{d_{x_i}=k \land d_{x_{i+1}}=k'\}} + 1_{\{d_{x_i}=k' \land d_{x_{i+1}}=k\}}).
\end{align*}
Finally, the hybrid unbiased estimator $\hat{P}(k, k')$ is defined with $\hat{\bar{k}}$ as a threshold:
\begin{align*}
\hat{P}(k, k') \triangleq
\begin{cases}
\hat{P}_{\text{IE}}(k, k') & (\text{if } k + k' \geq 2 \hat{\bar{k}}), \\
\hat{P}_{\text{TE}}(k, k') & (\text{if } k + k' < 2 \hat{\bar{k}}).
\end{cases}
\end{align*}
The original paper \cite{gjoka_2_5_k} did not prove that $\hat{P}(k, k')$ is an unbiased estimator of $P(k, k')$.
Therefore, we prove that in Appendix \ref{appendix:A}.

An unbiased estimator of the degree-dependent clustering coefficient \cite{katzir_nodecc} is given by $\hat{\bar{c}}(k) \triangleq \Phi_{\bar{c}}(k)/\Phi(k)$, where we define 
\begin{align*}
\Phi_{\bar{c}}(k) = \frac{1}{(k-1)(r-2)} \sum_{i=2}^{r-1} 1_{\{d_{x_i} = k\}} A_{x_{i-1} x_{i+1}}.
\end{align*}

\section{Proposed Method} \label{section:4}

\begin{figure*}[p]
        \begin{center}
          \includegraphics[scale=0.15]{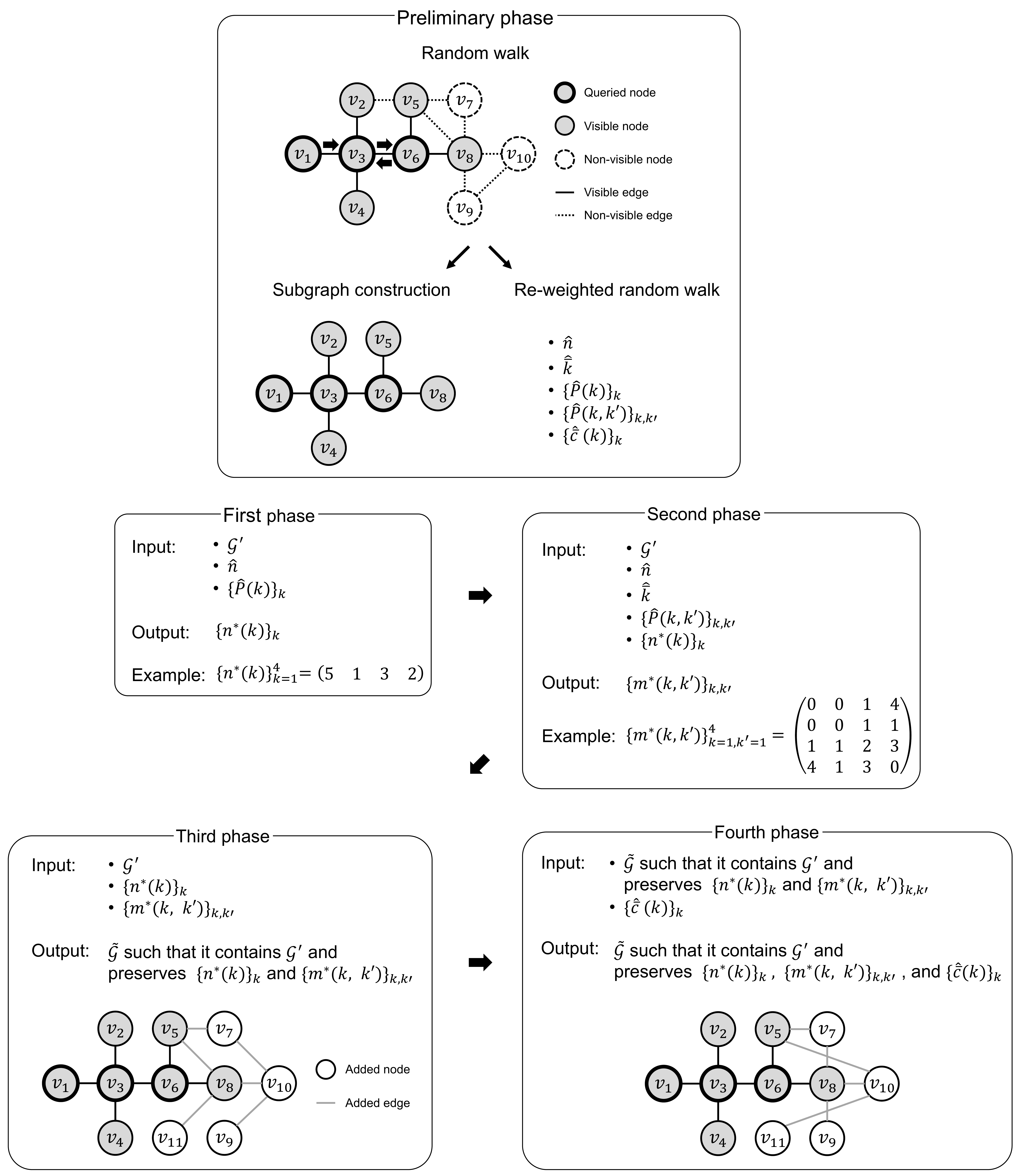}
        \end{center}
      \caption{Workflow of the proposed method. $\mathcal{G}'$ represents the subgraph obtained by a random walk; $\hat{n}$ represents the estimate of the number of nodes; $\hat{\bar{k}}$ represents the estimate of the average degree; $\{\hat{P}(k)\}_k$ represents the estimate of the degree distribution; $\{n^*(k)\}_k$ represents the target degree vector; $\{\hat{P}(k, k')\}_{k, k'}$ represents the estimate of the joint degree distribution; $\{m^*(k, k')\}_k$ represents the target joint degree matrix; $\{\hat{\bar{c}}(k)\}_k$ represents the estimate of the degree-dependent clustering coefficient; and $\tilde{\mathcal{G}}$ represents the graph to be generated by the proposed method.}
      \label{fig:2}
\end{figure*}

\subsection{Overview}
In this section, we propose a method for restoring the original graph $\mathcal{G}$ based on the subgraph $\mathcal{G}'$ and the estimates of five local structural properties (i.e., the number of nodes $\hat{n}$, average degree $\hat{\bar{k}}$, degree distribution $\{\hat{P}(k)\}_k$, joint degree distribution $\{\hat{P}(k, k')\}_{k, k'}$, and degree-dependent clustering coefficient $\{\hat{\bar{c}}(k)\}_k$).
Our idea is to add nodes and edges to the subgraph to generate a graph that preserves these estimates of local structural properties.
We intend to reproduce the global structural properties of the original graph by preserving local structural properties, as in the underlying idea of the $dK$-series \cite{mahadevan, orsini, gjoka_2_5_k}. 
Furthermore, we intend to reproduce the structural properties and the visual representation of the original graph more accurately by preserving the structure of the subgraph.

The proposed method consists of four phases (see also Fig. \ref{fig:2}).
We denote the graph to be generated by the proposed method as $\tilde{\mathcal{G}}$ throughout this section.
In the first phase, we construct the target degree vector, as denoted by $\{n^*(k)\}_k$ (Section \ref{section:4.B}).
This vector determines the number of nodes with degree $k$ in $\tilde{\mathcal{G}}$.
We construct the target degree vector based on the subgraph $\mathcal{G}'$ and the estimates $\hat{n}$ and $\{\hat{P}(k)\}_k$.
In the second phase, we construct the target joint degree matrix, as denoted by $\{m^*(k, k')\}_{k, k'}$ (Section \ref{section:4.C}).
This matrix determines the number of edges between nodes with degree $k$ and nodes with degree $k'$ in $\tilde{\mathcal{G}}$.
We construct the target joint degree matrix based on the subgraph $\tilde{\mathcal{G}}$, the estimates $\hat{n}$, $\hat{\bar{k}}$, and $\{\hat{P}(k, k')\}_{k, k'}$, and the target degree vector $\{n^*(k)\}_k$.
In the third phase, we add nodes and edges to the subgraph to ensure that the generated graph $\tilde{\mathcal{G}}$ preserves $\{n^*(k)\}_k$ and $\{m^*(k, k')\}_{k, k'}$ (Section \ref{section:4.D}).
In the fourth phase, we repeatedly rewire edges in the generated graph $\tilde{\mathcal{G}}$ so that $\tilde{\mathcal{G}}$ also preserves the estimate $\{\hat{\bar{c}}(k)\}_k$ (Section \ref{section:4.E}).
In the following sections, we describe each phase of the proposed method in detail.

\subsection{Constructing the target degree vector} \label{section:4.B}
In the first phase, we construct the target degree vector based on the subgraph $\mathcal{G}'$ and the estimates of the number of nodes $\hat{n}$ and the degree distribution $\hat{P}(k)$.
The target degree vector, as denoted by $\{n^*(k)\}_k$, determines the number of nodes with degree $k$ in the graph to be generated $\tilde{\mathcal{G}}$.

We denote by $k^*_{\text{max}}$ the target maximum degree of the graph to be generated, $\tilde{\mathcal{G}}$.
In general, a target degree vector, $\{n^*(k)\}_k$, needs to satisfy the following two conditions to realize a graph that preserves it \cite{newman_random}:
\begin{enumerate}[(DV-1)]
\setlength{\leftskip}{0.3cm}
\item $n^*(k)$ is a nonnegative integer for each $k = 1, \ldots, k^*_{\text{max}}$.
\item $\sum_{k=1}^{k^*_{\text{max}}} k n^*(k)$ is an even number.
\end{enumerate}

However, the immediate estimate of the number of nodes with degree $k$ obtained by $\hat{n}(k) = \hat{n} \hat{P}(k)$ (see Eq. \eqref{eq:2}) typically does not satisfy these realization conditions.
For example, $\hat{n}(k)$ is not typically an integer for each degree $k$.
Therefore, we construct $\{n^*(k)\}_k$, which satisfies the realization conditions while minimizing the error of $\{n^*(k)\}_k$ relative to the original estimate $\{\hat{n}(k)\}_k$.

\subsubsection{Initialization step}
We initialize $n^*(k)$ for each degree $k$ using $\hat{n}$ and $\hat{P}(k)$ such that $\{n^*(k)\}_k$ satisfies condition (DV-1).
Let $\text{NearInt}(a)$ denote a function that returns the nearest integer to real value $a$ and $\text{max}(b, c)$ denote a function that returns the larger of the two integers $b$ and $c$. 

First, we set the target maximum degree $k^*_{\text{max}}$ as the larger value between the maximum degree $k$ such that $\hat{P}(k) > 0$ and the maximum degree of the node in the subgraph $\mathcal{G}'$.
Then, for each degree $k =1, \ldots, k^*_{\text{max}}$, we set
\begin{align*}
n^*(k) = 
\begin{cases}
\text{max}(\text{NearInt}(\hat{n} \hat{P}(k)), 1) & \text{if }\hat{P}(k) > 0, \\
0 & \text{if } \hat{P}(k) = 0.
\end{cases}
\end{align*}
Note that we initialize $n^*(k)$ with a positive integer for each degree $k$ such that $\hat{P}(k) > 0$.
For example, if $\hat{n}\hat{P}(1) = 0.1$, we set $n^*(1) = 1$ and not $n^*(1) = 0$.
This is because if we obtain a positive estimate $\hat{P}(k) > 0$ then there must be at least one node with degree $k$ in the original graph $\mathcal{G}$ based on the definition of $\hat{P}(k)$.

\begin{algorithm}[t]
\caption{Adjust the target degree vector to ensure that it satisfies condition (DV-2).}
\label{alg:1}
\begin{algorithmic}[1]
\REQUIRE Estimates: $\hat{n}$ and $\{\hat{P}(k)\}_{k}$. \\
\REQUIRE Target maximum degree: $k_{\text{max}}^*$. \\
\REQUIRE Target degree vector: $\{n^*(k)\}_k$. \\
\IF{$\sum_{k=1}^{k^*_{\text{max}}} k n^*(k)$ is an odd number}
\STATE{Select degree $k$ such that $k$ is an odd number and $\Delta_{+}(k)$ is the smallest.}
\STATE{$n(k) \leftarrow n(k) + 1$.}
\ENDIF
\RETURN{$\{n^*(k)\}_k$.}
\end{algorithmic}
\end{algorithm}

\subsubsection{Adjustment step}
Then, we adjust the target degree vector $\{n^*(k)\}_k$ such that $\{n^*(k)\}_k$ satisfies condition (DV-2) as follows (see also Algorithm \ref{alg:1}). 
If and only if the sum of degrees, i.e., $\sum_{k=1}^{k^*_{\text{max}}} k n^*(k)$, is an odd number, we increase $n^*(k)$ by one for degree $k\ (1 \leq k \leq k^*_{\text{max}})$ such that $k$ is an odd number and the increase in the error of $n^*(k)$ relative to the original estimate $\hat{n}(k) = \hat{n} \hat{P}(k)$ upon increasing $n^*(k)$ by one, denoted by $\Delta_{+}(k)$, is the smallest value.
We define $\Delta_{+}(k)$ as
\begin{align*}
\Delta_{+}(k) = 
\begin{cases}
\frac{|\hat{n}(k) - (n^*(k)+1)|}{\hat{n}(k)} - \frac{|\hat{n}(k) - n^*(k)|}{\hat{n}(k)} & \text{if }\hat{P}(k) > 0, \\
\infty & \text{if } \hat{P}(k) = 0.
\end{cases}
\end{align*}
If there are two or more candidates for degree $k$ with the same increase $\Delta_{+}(k)$, we choose the smallest degree $k$ to minimize the increase in the number of edges in the graph to be generated, i.e., $\sum_{k=1}^{k_{\text{max}}^*} k n^*(k)$.
The target degree vector does not break condition (DV-1) because the adjustment step comprises only increasing $n^*(k)$ for some degree $k$.

\subsubsection{Modification step}
In general, to generate a graph that preserves a given target degree vector, we assign a target degree to each node, i.e., the degree of each node in the generated graph \cite{newman_random, mahadevan}.
Therefore, we next assign the target degree, denoted by $d_i^*$, of each node $v_i'$ in the subgraph $\mathcal{G}'$ whose target degree is constrained by the degree of the subgraph.
In parallel with this assignment process, we modify the target degree vector to ensure that it also satisfies another condition to realize a generated graph $\tilde{\mathcal{G}}$ that contains the subgraph $\mathcal{G}'$.
Specifically, the target number of nodes with degree $k$, i.e., $n^*(k)$, needs to be no less than the number of nodes with the target degree $k$ in the subgraph, which is defined as $n'(k) = \sum_{v_i' \in \mathcal{V}'} 1_{\{d_i^* = k\}}$, for each degree $k$:
\begin{enumerate}[(DV-3)]
\setlength{\leftskip}{0.3cm}
\item $n^*(k) \geq n'(k)$ for each degree $k =1, \ldots, k^*_{\text{max}}$.
\end{enumerate}
We modify the target degree vector to ensure that it satisfies all conditions (i.e., (DV-1), (DV-2), and (DV-3)) while minimizing the error of $\{n^*(k)\}_k$ relative to the original estimate $\{\hat{n}(k)\}_k$.

Before we assign the target degree of each node $v_i'$ in the subgraph $\mathcal{G}'$, we clarify the relationship between the degree $d_i'$ of $v_i'$ in $\mathcal{G}'$ and the degree $d_i$ of $v_i'$ in the original graph $\mathcal{G}$:
\begin{lemma}
\label{lemma:1}
For each node in the subgraph $v_i' \in \mathcal{V}'$, the degree $d_i'$ in $\mathcal{G}'$ and the degree $d_i$ in $\mathcal{G}$ satisfy 
\begin{align*}
d_i &= d'_i\ \ \ {\rm if}\ v_i' \in \mathcal{V}'_{\text{qry}}, \\
d_i &\geq d'_i\ \ \ {\rm if}\ v_i' \in \mathcal{V}'_{\text{vis}}.
\end{align*}
\end{lemma}
\begin{IEEEproof}
The first equation holds because all edges incident to a queried node are contained in $\mathcal{G}'$ based on the problem definition.
The second inequality holds because a visible node is connected only to queried neighbors in $\mathcal{G}'$.
\end{IEEEproof}

\begin{algorithm}[t]
\caption{Modify the target degree vector to ensure that it satisfies condition (DV-3).}
\label{alg:2}
\begin{algorithmic}[1]
\REQUIRE Subgraph: $\mathcal{G}' =(\mathcal{V}', \mathcal{E}')$. \\
\REQUIRE Estimates: $\hat{n}$ and $\{\hat{P}(k)\}_{k}$. \\
\REQUIRE Target maximum degree: $k_{\text{max}}^*$. \\
\REQUIRE Target degree vector: $\{n^*(k)\}_k$. \\
\STATE{Calculate degree $d_i'$ of each node in the subgraph $v_i' \in \mathcal{V}'$.}
\FOR{each $v_i' \in \mathcal{V}'_{\text{qry}}$ in arbitrary order}
\STATE{$d_i^* \leftarrow d_i'$.}
\ENDFOR
\STATE{Calculate the present $n'(k)$ for each $k=1, \ldots, k^*_{\text{max}}$.}
\FOR{$k=1, \ldots, k^*_{\text{max}}$}
\STATE{$n^*(k) \leftarrow \text{max}(n^*(k), n'(k))$.}
\ENDFOR
\FOR{each $v_i' \in \mathcal{V}'_{\text{vis}}$ in decreasing order of $d_i'$}
\STATE{Construct the degree sequence $\mathcal{D}_{\text{seq}}(i)$.}
\IF{$\mathcal{D}_{\text{seq}}(i)$ is not empty}
\STATE{Select degree $k$ uniformly randomly from $\mathcal{D}_{\text{seq}}(i)$.}
\ELSE
\STATE{Select degree $k$ such that $d_i' \leq k \leq k^*_{\text{max}}$ and $\Delta_{+}(k)$ is the smallest.}
\ENDIF
\STATE{$d_i^* \leftarrow k$.}
\STATE{$n'(k) \leftarrow n'(k) + 1$.}
\STATE{$n^*(k) \leftarrow \text{max}(n^*(k), n'(k))$.}
\ENDFOR
\RETURN{$\{n^*(k)\}_k$.}
\end{algorithmic}
\end{algorithm}

We modify the target degree vector as follows (see also Algorithm \ref{alg:2}).
First, for each queried node $v_i' \in \mathcal{V}_{\text{qry}}'$ in arbitrary order, we assign a target degree $d_i^* = d_i'$ that is the same as the degree of the node in the subgraph $\mathcal{G}'$, according to Lemma \ref{lemma:1} (lines 2--3 in Algorithm \ref{alg:2}).
Second, we calculate the present number of nodes with target degree $k$ in the subgraph, i.e., $n'(k)$, for each degree $k=1, \ldots, k^*_{\text{max}}$ (line 4 in Algorithm \ref{alg:2}).
Then, we modify the target number $n^*(k)$ to $n^*(k) = n'(k)$ if and only if $n^*(k) < n'(k)$ for each degree $k=1, \ldots, k^*_{\text{max}}$ to satisfy condition (DV-3) (lines 5--6 in Algorithm \ref{alg:2}).

Next, for each visible node $v_i' \in \mathcal{V}'_{\text{vis}}$, we assign the target degree $d_i^*$ such that $d_i^* \geq d_i'$, according to Lemma \ref{lemma:1}.
For this purpose, we first select the visible node $v_i' \in \mathcal{V}'_{\text{vis}}$ that is not assigned the target degree and has the largest degree in the subgraph.
Second, we construct a sequence of target degrees that can be assigned to $v_i'$, as denoted by $\mathcal{D}_{\text{seq}}(i)$, in which degree $k$ appears $n^*(k) - n'(k)$ times for each $k=d_i', \ldots, k^*_{\text{max}}$ (line 8 in Algorithm \ref{alg:2}).
If $\mathcal{D}_{\text{seq}}(i)$ is not empty, we choose degree $k$ uniformly and randomly from $\mathcal{D}_{\text{seq}}(i)$ (lines 9--10 in Algorithm \ref{alg:2}).
Otherwise, we select degree $k$ such that $d_i' \leq k \leq k^*_{\text{max}}$ and the increase in the error $\Delta_{+}(k)$ is the smallest (lines 11--12 in Algorithm \ref{alg:2}).
If two or more candidates exist for degree $k$ with the same increase $\Delta_{+}(k)$, we choose the smallest.
Then, we assign the target degree of $v_i'$ as $d_i^* = k$ and increase $n'(k)$ by one (lines 13--14 in Algorithm \ref{alg:2}).
We modify the target number $n^*(k)$ to $n^*(k) = n'(k)$ if and only if $n^*(k) < n'(k)$ (line 15 in Algorithm \ref{alg:2}).
We continue this procedure until we assign a target degree to all visible nodes.

Note that we assign target degrees to visible nodes in decreasing order of the degrees in the subgraph.
This is because a node with a larger degree in the subgraph tends to have fewer candidate target degrees in social graphs with heavy-tailed degree distributions \cite{ahn, mislove, gjoka_practical, gjoka_walking, kwak}.

The target degree vector, $\{n^*(k)\}_k$, does not break condition (DV-1) if we execute the modification algorithm on the target degree vector.
This is because the algorithm comprises only increasing $n^*(k)$ values for multiple degrees of $k$.
On the other hand, this modification step may make $\{n^*(k)\}_k$ break condition (DV-2).
In this case, we perform the adjustment process (Algorithm \ref{alg:1}) again.
If we execute the adjustment algorithm on the target degree vector, $\{n^*(k)\}_k$ does not break conditions (DV-1) and (DV-3).
This is because the algorithm comprises only increasing $n^*(k)$ for some degree $k$.
Therefore, $\{n^*(k)\}_k$ finally satisfies all conditions, i.e., (DV-1), (DV-2), and (DV-3).

\subsection{Constructing the target joint degree matrix} \label{section:4.C}
In the second phase, we construct the target joint degree matrix based on the subgraph $\mathcal{G}'$, the target degree vector $\{n^*(k)\}_k$, and the estimates of the number of nodes $\hat{n}$, the average degree $\hat{\bar{k}}$, and the joint degree distribution $\hat{P}(k, k')$.
The target joint degree matrix, as denoted by $\{m^*(k, k')\}_{k, k'}$, determines the number of edges between nodes with degree $k$ and nodes with degree $k'$ in the graph to be generated $\tilde{\mathcal{G}}$.

As in the case of constructing the target degree vector, the target joint degree matrix, $\{m^*(k, k')\}_{k, k'}$, needs to satisfy the following three conditions to realize a graph that preserves it:
\begin{enumerate}[(JDM-1)]
\setlength{\leftskip}{0.3cm}
\item $m^*(k, k')$ is a nonnegative integer for each degree $k=1, \ldots, k^*_{\text{max}}$ and $k'=1, \ldots, k^*_{\text{max}}$.
\item $m^*(k, k') = m^*(k', k)$ for each degree $k=1, \ldots, k^*_{\text{max}}$ and each $k=1, \ldots, k^*_{\text{max}}$ such that $k \neq k'$.
\item $\sum_{k'=1}^{k^*_{\text{max}}} \mu(k, k') m^*(k, k') = k n^*(k)$ for each degree $k=1, \ldots, k^*_{\text{max}}$.
\end{enumerate}
These conditions are obtained by relaxing the conditions required to realize a graph that preserves $\{m^*(k, k')\}_{k, k'}$ and contains no multiple edges or self-loops \cite{stanton}.

However, as in the case of constructing the target degree vector, the immediate estimate of the number of edges between nodes with degree $k$ and nodes with degree $k'$ obtained by $\hat{m}(k, k') = \hat{n} \hat{\bar{k}} \hat{P}(k, k')/\mu(k,k')$ (see Eqs. \eqref{eq:1} and \eqref{eq:3}) typically does not satisfy the realization conditions. 
Therefore, we construct $m^*(k, k')$ for each $k$ and $k'$ that satisfies the realization conditions while minimizing the error of $\{m^*(k, k')\}_{k, k'}$ relative to the original estimate $\{\hat{m}(k, k')\}_{k, k'}$.

\subsubsection{Initialization step}
First, we initialize $m^*(k, k')$ for each degree $k$ and $k'$ using $\hat{n}$, $\hat{\bar{k}}$, and $\hat{P}(k, k')$ such that it satisfies conditions (JDM-1) and (JDM-2).
Specifically, for each degree $k=1, \ldots, k^*_{\text{max}}$ and $k'=1, \ldots, k^*_{\text{max}}$, we set
\begin{align*}
&m^*(k, k') = \\
&\begin{cases}
\text{max}(\text{NearInt}(\hat{n} \hat{\bar{k}} \hat{P}(k, k')/\mu(k,k'), 1) & \text{if }\hat{P}(k, k') > 0, \\
0 & \text{if }\hat{P}(k, k') = 0.
\end{cases}
\end{align*}
It holds that $m^*(k, k') = m^*(k', k)$ because $\hat{P}(k, k') = \hat{P}(k', k)$ holds for $k \neq k'$.
Note that we initialize $m^*(k, k')$ as a positive integer for each degree $k$ and $k'$ such that $\hat{P}(k, k') > 0$.
This is because if we obtain a positive estimate $\hat{P}(k, k') > 0$ then there must be at least one edge between nodes with degree $k$ and nodes with degree $k'$ in the original graph $\mathcal{G}$ based on the definition of $\hat{P}(k, k')$.

\subsubsection{Adjustment step}
Then, we adjust $m^*(k, k')$ for each $k$ and $k'$ to ensure that it satisfies condition (JDM-3)  (see also Algorithm \ref{alg:3}).
We denote the present sum of $m^*(k, k')$ for degree $k' = 1, \ldots, k_{\text{max}}^*$ as $s(k) = \sum_{k'=1}^{k^*_{\text{max}}} \mu(k,k') m^*(k, k')$.
We also denote the target sum for degree $k$ as $s^*(k) = k n^*(k)$.
We denote the set of degrees $k$ by which we adjust the present sum $s(k)$ as $\mathcal{D}$.

For each degree $k \in \mathcal{D}$, we repeatedly increase or decrease $m^*(k, k')$ by one for multiple degrees $k'$ until the present sum $s(k)$ is equal to the target sum $s^*(k)$.
We define
\begin{align*}
\mathcal{D} = \{k \mid 1 \leq k \leq k^*_{\text{max}} \land s(k) \neq s^*(k)\} \cup \{1\}.
\end{align*}
We include degree $k=1$ in the set $\mathcal{D}$ to enable us to finely adjust the target joint degree matrix to ensure that it satisfies condition (JDM-3).
We adjust $n^*(k)$ if and only if $s(k)$ cannot reach $s^*(k)$ only with the adjustment of $m^*(k, k')$ for multiple degrees $k'$.

\begin{algorithm}[t]
\caption{Adjust the target joint degree matrix to ensure that it satisfies condition (JDM-3).}
\label{alg:3}
\begin{algorithmic}[1]
\REQUIRE Estimates: $\hat{n}$, $\hat{\bar{k}}$, and $\{\hat{P}(k, k')\}_{k, k'}$. \\
\REQUIRE Target maximum degree: $k_{\text{max}}^*$. \\
\REQUIRE Target degree vector: $\{n^*(k)\}_k$. \\
\REQUIRE Target joint degree matrix: $\{m^*(k, k')\}_{k, k'}$.\\
\REQUIRE Lower limits: $\{m_{\text{min}}(k, k')\}_{k, k'}$.
\FOR{each $k \in \mathcal{D}$ in decreasing order of $k$}
\IF{$k = 1$ and $|s(1) - s^*(1)|$ is an odd number} \label{line:2}
\STATE $n^*(1) \leftarrow n^*(1) + 1$. \label{line:3}
\ENDIF
\WHILE{$s(k) \neq s^*(k)$}
\IF{$s(k) < s^*(k)$} \label{line:5}
\STATE Select degree $k' \in \mathcal{D}'_{+}(k)$ with the smallest $\Delta_{+}(k, k')$.
\STATE $m^*(k, k') \leftarrow m^*(k, k') + 1$.
\IF{$k \neq k'$}
\STATE $m^*(k', k) \leftarrow m^*(k', k) + 1$. \label{line:9}
\ENDIF
\ELSE \label{line:10}
\IF{$\mathcal{D}_{-}'(k)$ is not empty}
\STATE Select degree $k' \in \mathcal{D}_{-}'(k)$ with the smallest $\Delta_{-}(k, k')$.
\STATE $m^*(k, k') \leftarrow m^*(k, k') - 1$.
\IF{$k \neq k'$}
\STATE $m^*(k', k) \leftarrow m^*(k', k) - 1$.
\ENDIF
\ELSE
\IF{$k = 1$}
\STATE $n^*(1) \leftarrow n^*(1) + 2$.
\ELSE
\STATE $n^*(k) \leftarrow n^*(k) + 1$. \label{line:20}
\ENDIF
\ENDIF
\ENDIF
\ENDWHILE
\ENDFOR
\RETURN{$\{m^*(k, k')\}_{k, k'}$.}
\end{algorithmic}
\end{algorithm}

We impose three constraints in adjusting the target joint degree matrix.
First, we ensure that $m^*(k, k')$ is not less than the input lower limit denoted by $m_{\text{min}}(k, k')$ for each $k$ and $k'$.
We assume that $m_{\text{min}}(k, k') \geq 0$.
The first constraint prevents $m^*(k, k')$ from violating condition (JDM-1) and is also used in the modification step, which is described in the next section.
Second, if we increase (decrease) $m^*(k, k')$ by one for $k' \neq k$, we also increase (decrease) $m^*(k', k)$ by one.
The second constraint prevents $m^*(k, k')$ and $m^*(k', k)$ such that $k \neq k'$ from violating condition (JDM-2).
The second constraint makes it difficult to adjust the target joint degree matrix because both present sums $s(k)$ and $s(k')$ change if we change $m^*(k, k')$ for $k' \neq k$.
We address this difficulty by imposing the following third constraint: when we attempt to adjust the present sum $s(k)$ for degree $k$, we do not change $m^*(k, k')$ for any degree $k'$ such that $s(k') = s^*(k')$ already holds true before adjusting the present sum $s(k)$.
The third constraint prevents the sum $s(k')$ for any degree $k'$ already been adjusted before adjusting the sum $s(k)$ from violating condition (JDM-3).

We adjust the present sum $s(k)$ for degree $k \in \mathcal{D}$ in descending order of $k$.
This ordering is based on the following two observations: (i) the later the adjustment order of $s(k)$ is, the fewer the elements of $m^*(k, k')$ that can be changed, and (ii) the smaller degree $k$ is, the fewer the edges that need to be added to make $s(k)$ equal to $s^*(k)$.
Accordingly, when we adjust $s(k)$, we are allowed to change only $m^*(k, k')$ for degree $k' \in \mathcal{D}$ such that $k' \leq k$.

When we adjust the present sum of degree 1, i.e., $s(1)$, we first need to ensure that the absolute difference between the present and target sums, i.e., $|s(1) - s^*(1)|$, is an even number. 
This reason is as follows. 
The sum $s(1)$ increases or decreases only by an even number because we are allowed to increase or decrease only $m^*(1, 1)$ due to our third constraint.
Thus, if $|s(1) - s^*(1)|$ is an odd number, $s(1)$ will not reach $s^*(1)$.
Therefore, in this case, we make the absolute difference an even number by increasing $n^*(1)$ by one (lines 2--3 in Algorithm \ref{alg:3}).

If $s(k) < s^*(k)$, we increase $m^*(k, k')$ by one for degree $k'$ (lines 5--9 in Algorithm \ref{alg:3}).
For the given $k$, we define the set of degrees $k'$ for which $m^*(k, k')$ is increased by one as 
\begin{align*}
\mathcal{D}'_{+}(k) = 
\begin{cases}
	\{k' \mid k' \in \mathcal{D} \land k' \leq k \} & \text{if }s(k) \neq s^*(k) - 1, \\
	\{k' \mid k' \in \mathcal{D} \land k' < k \} & \text{if }s(k) = s^*(k) - 1.
\end{cases}
\end{align*}
We exclude degree $k$ if $s(k) = s^*(k) - 1$ to avoid increasing $s(k)$ by two, where we recall that $s(k)$ is increased by two if we increase $m^*(k, k)$ by one because of the factor $\mu(k, k) = 2$.
The set $\mathcal{D}'_{+}(k)$ always contains at least one degree $k'$.
This is because if $k > 1$, the set $\mathcal{D}'_{+}(k)$ contains at least degree $k'=1$; otherwise, it holds that $\mathcal{D}'_{\text{inc}}(1) = \{1\}$ because our adjustment algorithm maintains $|s^*(1) - s(1)|$ as an even number.
Then, we increase $m^*(k, k')$ by one for degree $k' \in \mathcal{D}'_{+}(k)$ such that the increase in the error of $m^*(k, k')$ relative to the original estimate $\hat{m}(k, k') = \hat{n} \hat{\bar{k}} \hat{P}(k, k')/\mu(k,k')$ upon increasing $m^*(k, k')$ by one, as denoted by $\Delta_{+}(k, k')$, is the smallest.
We define $\Delta_{+}(k, k')$ as
\begin{align*}
&\Delta_{+}(k, k') = \\
&\begin{cases}
\frac{|\hat{m}(k, k') - (m^*(k, k')+1)|}{\hat{m}(k, k')} - \frac{|\hat{m}(k, k') - m^*(k, k')|}{\hat{m}(k, k')} & \text{if }\hat{P}(k, k') > 0, \\
 \infty & \text{if }\hat{P}(k, k') = 0.
\end{cases}
\end{align*}
If two or more candidates for degree $k'$ exist with the same increase $\Delta_{+}(k, k')$, we uniformly and randomly choose from among the $k'$ values. 
This random selection is based on our preliminary observation.

If $s(k) > s^*(k)$, we attempt to decrease $m^*(k, k')$ by one for degree $k'$ (lines 10--20 in Algorithm \ref{alg:3}).
We strictly adhere to the lower limit of $m^*(k, k')$, i.e., $m_{\text{min}}(k, k')$, for each $k$ and $k'$.
Unless we state otherwise, we set the lower limit as
\begin{align*}
m_{\text{min}}(k, k') = 0
\end{align*}
for any $k$ and $k'$.
For the given $k$, we define the set of degrees $k'$ for which $m^*(k, k')$ is decreased by one as
\begin{align*}
&\mathcal{D}_{-}'(k) = \\
&\begin{cases}
	\{k' \mid k' \in \mathcal{D} \land k' \leq k \land m^*(k, k') > m_{\text{min}}(k, k')\} & \\
	\ \ \ \ \ \ \ \ \ \ \ \ \ \ \ \ \ \ \ \ \ \ \ \ \ \ \ \ \ \ \ \ \ \ \ \ \ \ \text{if }s(k) \neq s^*(k) + 1, \\
	\{k' \mid k' \in \mathcal{D} \land k' < k \land m^*(k, k') > m_{\text{min}}(k, k')\} & \\
	\ \ \ \ \ \ \ \ \ \ \ \ \ \ \ \ \ \ \ \ \ \ \ \ \ \ \ \ \ \ \ \ \ \ \ \ \ \ \text{if }s(k) = s^*(k) + 1.
\end{cases}
\end{align*}
If $\mathcal{D}_{-}'(k)$ is not an empty set, we decrease $m^*(k, k')$ by one for degree $k' \in \mathcal{D}_{-}'(k)$ such that the increase in the error of $m^*(k, k')$ relative to the original estimate $\hat{m}(k, k')$ upon decreasing $m^*(k, k')$ by one, as denoted by $\Delta_{-}(k, k')$, is the smallest.
We define $\Delta_{-}(k, k')$ as
\begin{align*}
&\Delta_{-}(k, k') = \\
&\begin{cases}
 \frac{|\hat{m}(k, k') - (m^*(k, k')-1)|}{\hat{m}(k, k')} - \frac{|\hat{m}(k, k') - m^*(k, k')|}{\hat{m}(k, k')} & \text{if }\hat{P}(k, k') > 0, \\
 \infty & \text{if }\hat{P}(k, k') = 0.
\end{cases}
\end{align*}
If two or more candidates for degree $k'$ exist, we uniformly and randomly choose $k'$ between among the $k'$ values.

The set $\mathcal{D}'_{-} (k)$ may be an empty set due to the constraint of the lower limits.
In this case, we increase the target sum $s^*(k)$ to shift the adjustment process toward the adjustment in which we increase the present sum $s(k)$.
Specifically, if $k > 1$, we increase $s^*(k)$ by $k$ by increasing $n^*(k)$ by one; otherwise, we increase $s^*(1)$ by two while maintaining $|s^*(1) - s(1)|$ as an even number by increasing $n^*(1)$ by two (lines 16--20 in Algorithm \ref{alg:3}).

For any degree $k \in \mathcal{D}$, the present sum $s(k)$ will reach the target sum $s^*(k)$ in a finite number of steps.
The reason is as follows.
If we increase $m(k, k')$ by one for degree $k' \neq k$, the present sum $s(k)$ increases by one.
If we increase $m(k, k)$ by one, the present sum $s(k)$ increases by two.
When $k > 1$, there is at least one degree $k'$ such that the sum $s(k)$ is increased by one because the set $\mathcal{D}'_{+}(k)$ always contains degree $k'=1$.
When $k = 1$, we are allowed to increase the sum $s(1)$ by only two because $\mathcal{D}'_{+}(1) = \{1\}$. 
However, we maintain $|s(1) - s^*(1)|$ as an even number throughout the process.
Therefore, $s(k)$ will reach $s^*(k)$ in the case of an adjustment by increasing.
In the case of an adjustment by decreasing, although there is the case in which the set $\mathcal{D}'_{-}(k)$ is an empty set, we recall that we shift the process toward the adjustment by increasing in that case.

\begin{algorithm}[t]
\caption{Modify the target joint degree matrix to ensure that it satisfies condition (JDM-4).}
\label{alg:4}
\begin{algorithmic}[1]
\REQUIRE Subgraph: $\mathcal{G}' =(\mathcal{V}', \mathcal{E}')$. \\
\REQUIRE Estimates: $\hat{n}$, $\hat{\bar{k}}$, and $\{\hat{P}(k, k')\}_{k, k'}$. \\
\REQUIRE Target maximum degree: $k_{\text{max}}^*$. \\
\REQUIRE Target degree of each node in the subgraph: $\{d_i^*\}_{v_i \in V^*}$ \\
\REQUIRE Target joint degree matrix: $\{m^*(k, k')\}_{k, k'}$. \\
\STATE{Calculate $m'(k, k')$ for each $k=1,\ldots,k^*_{\text{max}}$ and $k'=1,\ldots,k^*_{\text{max}}$.}
\FOR{$k_1 = 1, \ldots, k^*_{\text{max}}$}
\FOR{$k_2 = k_1, \ldots, k^*_{\text{max}}$}
\WHILE{$m^*(k_1, k_2) < m'(k_1, k_2)$}
\STATE{$m^*(k_1, k_2) \leftarrow m^*(k_1, k_2) + 1$.}
\IF{$k_1 \neq k_2$}
\STATE{$m^*(k_2, k_1) \leftarrow m^*(k_2, k_1) + 1$.}
\ENDIF
\IF{$\mathcal{D}''_{-}(k_1)$ is not empty}
\STATE{Select degree $k_3 \in \mathcal{D}''_{-}(k_1)$ with the smallest $\Delta_{-}(k_1, k_3)$.}
\STATE{$m^*(k_1, k_3) \leftarrow m^*(k_1, k_3) - 1$.}
\IF{$k_3 \neq k_1$}
\STATE{$m^*(k_3, k_1) \leftarrow m^*(k_3, k_1) - 1$.}
\ENDIF
\ENDIF
\IF{$\mathcal{D}''_{-}(k_2)$ is not empty}
\STATE{Select degree $k_4 \in \mathcal{D}''_{-}(k_2)$ with the smallest $\Delta_{-}(k_2, k_4)$.}
\STATE{$m^*(k_2, k_4) \leftarrow m^*(k_2, k_4) - 1$.}
\IF{$k_4 \neq k_2$}
\STATE{$m^*(k_4, k_2) \leftarrow m^*(k_4, k_2) - 1$.}
\ENDIF
\ENDIF
\IF{both degrees $k_3$ and $k_4$ have been found}
\STATE{$m^*(k_3, k_4) \leftarrow m^*(k_3, k_4) + 1$.}
\IF{$k_3 \neq k_4$}
\STATE{$m^*(k_4, k_3) \leftarrow m^*(k_4, k_3) + 1$.}
\ENDIF
\ENDIF
\ENDWHILE
\ENDFOR
\ENDFOR
\RETURN{$\{m^*(k, k')\}_{k, k'}$.}
\end{algorithmic}
\end{algorithm}

\subsubsection{Modification step}
Next, we modify the target joint degree matrix to ensure that it also satisfies the required condition for realizing a generated graph $\tilde{\mathcal{G}}$ that contains the subgraph $\mathcal{G}'$.
Specifically, the target number of nodes between nodes with degree $k$ and nodes with degree $k'$, i.e., $m^*(k, k')$, should be no less than the number of edges between nodes with target degree $k$ and nodes with target degree $k'$ in the subgraph, which is defined as $m'(k, k') = \sum_{(v_i', v_j') \in \mathcal{E}'} 1_{\{(d_i^* = k \land d_j^*=k') \lor (d_i^* = k' \land d_j^*=k)\}}$, for each $k$ and $k'$:
\begin{enumerate}[(JDM-4)]
\setlength{\leftskip}{0.3cm}
\item $m^*(k, k') \geq m'(k, k')$ for each $k =1, \ldots, k^*_{\text{max}}$ and $k' = 1, \ldots, k^*_{\text{max}}$.
\end{enumerate}
We modify the target joint degree matrix to ensure that it satisfies all conditions, i.e., (JDM-1), (JDM-2), (JDM-3), and (JDM-4), while minimizing the error of $\{m^*(k, k')\}_{k, k'}$ relative to the original estimate $\{\hat{m}(k, k')\}_{k, k'}$.

The basic idea behind our modification algorithm for the target joint degree matrix is as follows.
Suppose that $m^*(k_1, k_2) < m'(k_1, k_2)$ for a pair of degrees $k_1$ and $k_2$; hence, we need to modify $m^*(k_1, k_2)$ to ensure that $m^*(k_1, k_2) \geq m'(k_1, k_2)$ holds.
A simple modification involves forcibly increasing $m^*(k_1, k_2)$ to $m^*(k_1, k_2) = m'(k_1, k_2)$.
However, if one performs this forced increase on multiple pairs of $k_1$ and $k_2$, the sum $s(k)$ will violate condition (JDM-3) for multiple degrees $k$, and the target number of edges in the graph to be generated, i.e., $\sum_{k = 1}^{k_{\text{max}}^*} \sum_{k'= k}^{k_{\text{max}}^*} m^*(k, k')$, will cumulatively increase.
Therefore, if we increase $m^*(k_1, k_2)$ by one, we attempt to decrease $m^*(k_1, k_3)$ by one for degree $k_3$ such that $k_3 \neq k_2$ to ensure that both sums $s(k_1)$ and $s(k_2)$ are retained, minimizing the violation of condition (JDM-3) and the increase in the target number of edges.

Specifically, we modify the target joint degree matrix as follows (see also Algorithm \ref{alg:4}). 
For each pair of degrees $k_1$ and $k_2\ (1 \leq k_1 \leq k^*_{\text{max}},\ k_1 \leq k_2 \leq k^*_{\text{max}})$, we repeat the following procedure until $m^*(k_1, k_2)$ is not less than $m'(k_1, k_2)$.
First, we increase $m^*(k_1, k_2)$ by one (line 5 in Algorithm \ref{alg:4}).
If $k_1 \neq k_2$, we also increase $m^*(k_2, k_1)$ by one (lines 6--7 in Algorithm \ref{alg:4}).
Second, we attempt to find a degree $k_3$ such that $k_3 \neq k_2$ and $m^*(k_1, k_3)$ can be decreased by one to ensure that the sum $s(k_1)$ is retained.
We define a set of such degrees for the given degree $k$ as
\begin{align*}
&\mathcal{D}_{-}''(k) = \\
&\{k' \mid 1 \leq k' \leq k_{\text{max}}^* \land k' \neq k \land m^*(k, k') > m'(k, k')\}.
\end{align*}
If $\mathcal{D}''_{-}(k_1)$ is not empty, we select $k_3 \in \mathcal{D}''_{-}(k_1)$ with the smallest $\Delta_{-}(k_1, k_3)$ and decrease $m^*(k_1, k_3)$ by one (lines 8--10 in Algorithm \ref{alg:4}).
If there are two or more candidates for $k_3$ with the same increase $\Delta_{-}(k_1, k_3)$, we uniformly and randomly choose $k_3$ from among those candidates.
If $k_3 \neq k_1$, we also decrease $m^*(k_3, k_1)$ by one (lines 11--12 in Algorithm \ref{alg:4}).
Third, since $m^*(k_2, k_1)$ has been increased by one, we attempt to decrease $m(k_2, k_4)$ by one for degree $k_4$ such that $k_4 \neq k_2$ to ensure that the sum $s(k_2)$ is retained.
If the set $\mathcal{D}''_{-}(k_2)$ is not empty, we select the $k_4 \in \mathcal{D}''_{-}(k_2)$ with the smallest $\Delta_{-}(k_2, k_4)$ and decrease $m^*(k_2, k_4)$ by one (lines 13--15 in Algorithm \ref{alg:4}).
If two or more candidates for $k_4$ exist with the same increase $\Delta_{-}(k_2, k_4)$, we uniformly and randomly choose from among the $k_4$ values.
If $k_2 \neq k_4$, we also decrease $m^*(k_4, k_2)$ by one (lines 16--17 in Algorithm \ref{alg:4}).
Finally, if and only if both $k_3$ and $k_4$ have been found, we increase $m^*(k_3, k_4)$ by one and increase $m^*(k_4, k_3)$ by one if $k_3 \neq k_4$ to ensure that both sums $s(k_3)$ and $s(k_4)$ are retained (lines 18--21 in Algorithm \ref{alg:4}).

If we find either of degrees $k_3$ and $k_4$, $m^*(k_1, k_2)$ is increased by one while preserving the target number of edges, i.e., $\sum_{k = 1}^{k_{\text{max}}^*} \sum_{k'= k}^{k_{\text{max}}^*} m^*(k, k')$.
This is because $m^*(k_1, k_2)$ is increased by one, and either $m^*(k_1, k_3)$ or $m^*(k_2, k_4)$ is decreased by one. 
Furthermore, if both $k_3$ and $k_4$ have been found, $m^*(k_1, k_2)$ is increased by one while preserving the target number of edges and the sum $s(k)$ for any degree $k$ because all  sums $s(k_1), s(k_2), s(k_3),$ and $s(k_4)$ are retained.

The target joint degree matrix, i.e., $\{m^*(k, k')\}_{k, k'}$, does not break conditions (JDM-1) and (JDM-2) if we execute the modification algorithm on the target joint degree matrix.
On the other hand, $\{m^*(k, k')\}_{k, k'}$ may break condition (JDM-3). 
In this case, we again execute the adjustment algorithm on the target joint degree matrix (Algorithm \ref{alg:3}), with the lower limit $m_{\text{min}}(k, k')$ set as
\begin{align*}
m_{\text{min}}(k, k') = m'(k, k')
\end{align*}
for each degree $k$ and $k'$ to ensure that $\{m^*(k, k')\}_{k, k'}$ retains condition (JDM-4).
If we perform the adjustment algorithm again, $\{m^*(k, k')\}_{k, k'}$ still satisfies conditions (JDM-1), (JDM-2), and (JDM-4), and hence, it finally satisfies all conditions, i.e., (JDM-1), (JDM-2), (JDM-3), and (JDM-4).

\begin{algorithm}[t]                     
\caption{Construct a graph that preserves the target degree vector and the target joint degree matrix from the subgraph.}         
\label{alg:5}                          
\begin{algorithmic}[1]
\REQUIRE Subgraph: $\mathcal{G}'$. \\   
\REQUIRE Target degree vector: $\{n^*(k)\}_k$. \\
\REQUIRE Target joint degree matrix: $\{m^*(k, k')\}_{k, k'}$. \\
\STATE $\tilde{\mathcal{G}} \leftarrow \mathcal{G}'$.
\STATE Add $(\sum_{k=1}^{k_{\text{max}}^*} n^*(k)) - n'$ nodes to a set of nodes in $\tilde{\mathcal{G}}$.
\STATE Construct the degree sequence $\mathcal{D}_{\text{seq}}$ in which degree $k$ appears $n^*(k) - n'(k)$ times for $k = 1, \ldots, k^*_{\text{max}}$.
\STATE Randomly shuffle $\mathcal{D}_{\text{seq}}$.
\FOR{each added node $\tilde{v}_i \in \mathcal{V}_{\text{add}}$ in arbitrary order}
\STATE $k \leftarrow$ the last element in $\mathcal{D}_{\text{seq}}$.
\STATE Remove the last element in $\mathcal{D}_{\text{seq}}$.
\STATE $d_i^* \leftarrow k$.
\ENDFOR
\FOR{each node $\tilde{v}_i \in \mathcal{V}_{\text{qry}} \cup \mathcal{V}_{\text{vis}}$}
\STATE Attach $d_i^* - d_i'$ half-edges to $\tilde{v}_i$.
\ENDFOR
\FOR{each node $\tilde{v}_i \in \mathcal{V}_{\text{add}}$}
\STATE Attach $d_i^*$ half-edges to $\tilde{v}_i$.
\ENDFOR
\FOR{$k = 1, \ldots, k^*_{\text{max}}$}
\FOR{$k' = k, \ldots, k^*_{\text{max}}$}
\FOR{$i=1$ to $m^*(k, k') - m'(k, k')$}
\STATE Uniformly and randomly select a free half-edge of the nodes with the target degree $k$ and a free half-edge of the nodes with the target degree $k'$ and connect them.
\ENDFOR
\ENDFOR
\ENDFOR
\RETURN{$\tilde{\mathcal{G}}$.}
\end{algorithmic}
\end{algorithm}

\subsection{Adding nodes and edges to the subgraph} \label{section:4.D}
In the third phase, we add nodes and edges to the subgraph $\mathcal{G}'$ to ensure that the graph to be generated, $\tilde{\mathcal{G}}$, preserves the target degree vector $\{n^*(k)\}_k$ and the target joint degree matrix $\{m^*(k, k')\}_{k, k'}$.
It is trivial to construct a graph that preserves the given $\{n^*(k)\}_k$ and $\{m^*(k, k')\}_{k, k'}$ from an empty graph \cite{mahadevan, stanton, fosdick}.
We extend the existing construction procedure to the case of constructing the graph $\tilde{\mathcal{G}}$ that preserves $\{n^*(k)\}_k$ and $\{m^*(k, k')\}_{k, k'}$ from the subgraph $\mathcal{G}'$ (see also Algorithm \ref{alg:5}).

First, we set the graph $\tilde{\mathcal{G}}$ as the subgraph $\mathcal{G}'$ (line 1 in Algorithm \ref{alg:5}).
Second, we add $(\sum_{k=1}^{k_{\text{max}}^*} n^*(k)) - n'$ nodes to a set of nodes in the subgraph such that $\tilde{\mathcal{G}}$ contains $\sum_{k=1}^{k_{\text{max}}^*} n^*(k)$ nodes, where $n'$ is the number of nodes in the subgraph and $\sum_{k=1}^{k_{\text{max}}^*} n^*(k)$ is the target number of nodes in $\tilde{\mathcal{G}}$ (line 2 in Algorithm \ref{alg:5}).
We denote the set of added nodes as $\mathcal{V}_{\text{add}}$. 
We denote the set of nodes in $\tilde{\mathcal{G}}$ as $\tilde{\mathcal{V}}$.
It holds that $\tilde{\mathcal{V}}$ is a union of three disjoint sets, i.e., $\mathcal{V}_{\text{qry}}$, $\mathcal{V}_{\text{vis}}$, and $\mathcal{V}_{\text{add}}$.
Third, for each degree $k = 1, \ldots, k^*_{\text{max}}$ we arbitrarily assign a target degree $k$ to the $n^*(k) - n'(k)$ nodes that are not assigned a target degree in $\mathcal{V}_{\text{add}}$ (lines 3--8 in Algorithm \ref{alg:5}).
Note that $n^*(k) - n'(k) \geq 0$ always holds true because $\{n^*(k)\}_k$ satisfies condition (DV-3).
Fourth, we ensure that each node in the subgraph $\tilde{v}_i \in \mathcal{V}_{\text{qry}} \cup \mathcal{V}_{\text{vis}}$ has $d_i^* - d_i'$ half-edges, where $d^*_i$ is the target degree of $\tilde{v}_i$ and $d_i'$ is the degree of $\tilde{v}_i$ in the subgraph (lines 9--10 in Algorithm \ref{alg:5}).
Fifth, we ensure that each added node $\tilde{v}_i \in \mathcal{V}_{\text{add}}$ has $d_i^*$ half-edges (lines 11--12 in Algorithm \ref{alg:5}).
Finally, for each degree $k=1, \ldots, k^*_{\text{max}}$ and $k'=k, \ldots, k^*_{\text{max}}$, we repeat the following procedure $m^*(k, k') - m'(k, k')$ times: we randomly connect a free half-edge of the nodes with the target degree $k$ and a free half-edge of the nodes with the target degree $k'$ (lines 13--15 in Algorithm \ref{alg:5}).

\begin{algorithm}[t]                     
\caption{Rewire edges to ensure that the generated graph preserves the estimate of the degree-dependent clustering coefficient.}         
\label{alg:6}                          
\begin{algorithmic}[1]               
\REQUIRE Generated graph: $\tilde{\mathcal{G}} = (\tilde{\mathcal{V}}, \tilde{\mathcal{E}})$. \\
\REQUIRE Estimate: $\{\hat{\bar{c}}(k)\}_k$. \\
\REQUIRE Coefficient of the number of rewiring attempts: $R_{\text{C}}$. \\
\STATE $\tilde{\mathcal{E}}_{\text{rew}} \leftarrow$ a set of candidate edges to be rewired in $\tilde{\mathcal{G}}$.
\STATE $R \leftarrow R_{\text{C}} |\tilde{\mathcal{E}}_{\text{rew}}|$ // the number of rewiring attempts.
\STATE $\{\tilde{\bar{c}}(k)\}_k \leftarrow$ the present degree-dependent clustering coefficient of $\tilde{\mathcal{G}}$.
\STATE $D \leftarrow$ $L^1$ distance between $\{\tilde{\bar{c}}(k)\}_k)$ and $\{\hat{\bar{c}}(k)\}_k$.
\FOR{$r'=1$ to $R$}
\STATE{$(\tilde{v}_i, \tilde{v}_j), (\tilde{v}_a, \tilde{v}_b) \leftarrow$ random edge pair in $\tilde{\mathcal{E}}_{\text{rew}}$.}
\STATE{$\{\tilde{\bar{c}}_{\text{rew}}(k)\}_k \leftarrow$ degree-dependent clustering coefficient when the selected edge pair is rewired.} \label{ddcc_rew}
\STATE{$D_{\text{rew}} \leftarrow L^1$ distance between $\{\tilde{\bar{c}}_{\text{rew}}(k)\}_k$ and $\{\hat{\bar{c}}(k)\}_k$.}
\IF{$D_{\text{rew}} < D$}
\STATE{Remove edges $(\tilde{v}_i, \tilde{v}_j)$ and $(\tilde{v}_a, \tilde{v}_b)$.}
\STATE{Add edges $(\tilde{v}_i, \tilde{v}_b)$ and $(\tilde{v}_a, \tilde{v}_j)$.}
\STATE{Update $\tilde{\mathcal{E}}_{\text{rew}}$.}
\STATE{$D \leftarrow D_{\text{rew}}$.}
\ENDIF
\ENDFOR
\RETURN{$\tilde{\mathcal{G}}$.}
\end{algorithmic}
\end{algorithm}

\subsection{Rewiring edges in the generated graph} \label{section:4.E}
In general, it is practically difficult to generate a graph that exactly preserves a given degree-dependent clustering coefficient because the clustering coefficients of multiple nodes simultaneously change if an edge is added or removed \cite{orsini, tillman}.
In practice, one performs a large number of rewiring attempts of edges in a given graph to ensure that the graph approximately preserves the given degree-dependent clustering coefficient \cite{mahadevan, gjoka_2_5_k, orsini}. 

We perform the following process of rewiring edges in the generated graph $\tilde{\mathcal{G}}$ to ensure that $\tilde{\mathcal{G}}$ approximately preserves $\{\hat{\bar{c}}(k)\}_k$ (see also Algorithm \ref{alg:6}).
We first uniformly and randomly select an edge pair $(\tilde{v}_i, \tilde{v}_j) \in \tilde{\mathcal{E}}_{\text{rew}}$ and $(\tilde{v}_{i'}, \tilde{v}_{j'}) \in \tilde{\mathcal{E}}_{\text{rew}}$ such that the degrees of $\tilde{v}_i$ and $\tilde{v}_{i'}$ are equal, where $\tilde{\mathcal{E}}_{\text{rew}}$ is a set of candidate edges to be rewired.
We define $\tilde{\mathcal{E}}_{\text{rew}}$ as
\begin{align}
\tilde{\mathcal{E}}_{\text{rew}} = \tilde{\mathcal{E}} \setminus \mathcal{E}',
\label{eq:4}
\end{align}
where $\tilde{\mathcal{E}}$ represents a set of edges in $\tilde{\mathcal{G}}$.
Then, we replace $(\tilde{v}_i, \tilde{v}_j)$ and $(\tilde{v}_{i'}, \tilde{v}_{j'})$ with $(\tilde{v}_i, \tilde{v}_{j'})$ and $(\tilde{v}_{i'}, \tilde{v}_j)$ if and only if the normalized $L^1$ distance between the estimated and present degree-dependent clustering coefficients, denoted by $D$, decreases when we rewire the edges.
We define $D$ as
\begin{align*}
D = \frac{\sum_{k=1}^{k_{\text{max}}^*} |\tilde{\bar{c}}(k) - \hat{\bar{c}}(k)|}{\sum_{k=1}^{k_{\text{max}}^*} \hat{\bar{c}}(k)},
\end{align*}
where $\{\tilde{\bar{c}}(k)\}_k$ represents the present degree-dependent clustering coefficient of $\tilde{\mathcal{G}}$.
If the rewiring is accepted, we update the set $\tilde{\mathcal{E}}_{\text{rew}}$. 
We repeat this rewiring attempt a sufficiently large number of $R = R_{\text{C}} |\tilde{\mathcal{E}}_{\text{rew}}|$ times, where $R_{\text{C}}$ is a coefficient of the number of rewiring attempts.

The rewiring process exactly preserves both the degree vector $\{n^*(k)\}_k$ and the joint degree matrix $\{m^*(k, k')\}_{k, k'}$ of the generated graph $\tilde{\mathcal{G}}$ for the following two reasons.
First, the rewiring process preserves the degree of each node and hence preserves $\{n^*(k)\}_k$.
Second, the rewiring preserves $m^*(k, k')$ for each $k$ and $k'$ because the degrees of $\tilde{v}_i$ and $\tilde{v}_j$ are equal \cite{mahadevan, orsini, gjoka_2_5_k}.

We empirically require the rewiring of several hundred times the number of candidate edges \cite{gjoka_2_5_k, orsini}.
Thus, the exact recalculation of $\{\tilde{\bar{c}}(k)\}_k$ per rewiring attempt is not practical.
In practice, it is sufficient to update the difference in the number of triangles to which only nodes that are involved in the rewiring, i.e., $v_i, v_j, v_{i'}, v_{j'}$, and their neighbors, belong.
Updating the number of triangles to which a node involved in one rewiring attempt requires an average time of $O(\tilde{\bar{k}}^2)$, where $\tilde{\bar{k}}$ represents the average degree of $\tilde{\mathcal{G}}$.
In total, the rewiring algorithm requires an average time of $O(\tilde{\bar{k}}^2 R_{\text{C}} |\tilde{\mathcal{E}}_{\text{rew}}|)$.

The rewiring process exactly preserves the structure of the subgraph $\mathcal{G}'$ of $\tilde{\mathcal{G}}$ because we exclude the edges in the subgraph from the candidate edges to be rewired, as shown in Eq. \eqref{eq:4}.
In contrast, every edge in a given graph is a candidate edge to be rewired in Gjoka et al.'s  procedure (i.e., $\tilde{\mathcal{E}}_{\text{rew}} = \tilde{\mathcal{E}}$) \cite{gjoka_2_5_k}. 
This is because Gjoka et al.'s rewiring process does not use any structure of the subgraph sampled by a random walk.
Our rewiring procedure has two advantages over Gjoka et al.'s procedure because of the reduction in the number of candidate edges to be rewired: 
(i) our method is more likely to succeed in the rewiring of edges such that the generated graph approximately preserves $\{\hat{\bar{c}}(k)\}_k$, and
(ii) the proposed procedure reduces the rewiring time to $O(\tilde{\bar{k}}^2 R_{\text{C}} (|\tilde{\mathcal{E}}| - |\mathcal{E}'|))$ from $O(\tilde{\bar{k}}^2 R_{\text{C}} |\tilde{\mathcal{E}}|)$ in Gjoka et al.'s procedure.

\section{Experimental Design} \label{exp}
We evaluate the proposed method in terms of the accuracy of structural properties, the visual representation of generated graphs, and the generation time.
We conduct all experiments on a Linux server with an Intel Xeon E5-2698 (2.20 GHz) processor and 503 GB of main memory. 
All code is implemented in C++.
The datasets and source code used in our experiments are available at Ref. \cite{nakajima_data_code}.

\subsection{Datasets}
We use seven datasets of social graphs that are publicly available at Refs. \cite{nr, snap}. 
We preprocessed each dataset by first removing multiple edges and the directions of edges from the original graph and then by extracting the largest connected component.
Table \ref{datasets} lists the numbers of nodes and edges in all the graphs used in our experiments.

\begin{table}[t]
\caption{Datasets.}
\vspace{-3mm}
\label{datasets}
\begin{center}
	\begin{tabular}{l | c c c c}\hline
	Dataset & Number of nodes & Number of edges \\ \hline
	Anybeat \cite{nr} & 12,645 & 49,132 \\ 
	Brightkite \cite{nr} & 56,739 & 212,945 \\
	Epinions \cite{snap} & 75,877 & 405,739 \\
	Slashdot \cite{nr} & 77,360 & 469,180 \\
	Gowalla \cite{nr} & 196,591 & 950,327 \\
	Livemocha \cite{nr} & 104,103 & 2,193,083 \\
	YouTube \cite{snap} & 1,134,890 & 2,987,624 \\ 
	\hline
  	\end{tabular}
\end{center}
\vspace{-3mm}
\end{table}

\subsection{Structural properties of interest}
We focus on 12 structural properties of a given graph. 
\begin{enumerate}
\item Number of nodes, $n$.
\item Average degree, $\bar{k}$.
\item Degree distribution, $\{P(k)\}_k$.
\item Neighbor connectivity \cite{boccaletti}, as denoted by $\{\bar{k}_{\text{nn}}(k)\}_k$.
We define
\begin{align*}
\bar{k}_{\text{nn}}(k) = \frac{1}{n(k)} \sum_{i=1,\ d_i=k}^n \frac{1}{k} \sum_{j=1}^n A_{i, j} d_j
\end{align*}
for each $k$.
This property measures the average degree of neighbors of nodes with degree $k$, which is a coarse-grained version of the joint degree distribution \cite{mahadevan, orsini}.
\item Network clustering coefficient \cite{boccaletti}, as defined by 
\begin{align*}
\bar{c} = \frac{1}{n} \sum_{i=1}^n \frac{2 t_i}{d_i (d_i-1)}.
\end{align*}
\item Degree-dependent clustering coefficient, $\{\bar{c}(k)\}_k$.
\item Edgewise shared partner distribution \cite{hunter}, as denoted by $\{P(s)\}_s$.
We define
\begin{align*}
P(s) = \frac{1}{m} \sum_{(v_i, v_j) \in \mathcal{E}, i < j} 1_{\{\text{sp}(i, j) = s\}}
\end{align*}
for each $s$, where $\text{sp}(i, j) = \sum_{k=1,\ k \neq i,\ k \neq j}^n A_{i, k} A_{j, k}$.
This property measures the proportion of edges that have $s$ common neighbors.
\item Average shortest-path length, as denoted by $\bar{l}$. 
We define 
\begin{align*}
\bar{l} = \frac{2}{n(n-1)} \sum_{i=1}^{n-1} \sum_{j=i+1}^n l_{i, j},
\end{align*}
where $l_{i, j}$ denotes the shortest-path length between $v_i$ and $v_j$.
\item Shortest-path length distribution, as denoted by $\{P(l)\}_l$.
We define 
\begin{align*}
P(l) = \frac{2}{n(n-1)} \sum_{i=1}^{n-1} \sum_{j=i+1}^n 1_{\{l_{i, j} = l\}}.
\end{align*}
\item Diameter, which is the longest shortest-path length between two nodes and is denoted by $l_{\text{max}}$.
\item Degree-dependent betweenness centrality, as denoted by $\{\bar{b}(k)\}_k$.
We define 
\begin{align*}
\bar{b}(k) = \frac{1}{n(k)} \sum_{i=1,\ d_i=k}^n b_i,
\end{align*}
where $b_i$ is the betweenness centrality of $v_i$.
We define
\begin{align*}
b_i = \sum_{j=1,\ j \neq i}^n \sum_{k=1,\ k \neq i,\ k \neq j}^n \frac{\sigma_{j, k}(i)}{\sigma_{j, k}},
\end{align*}
where $\sigma_{j, k}(i)$ is the number of shortest paths between node $v_j$ and node $v_k$ that pass through node $v_i$ and $\sigma_{j, k}$ is the number of shortest paths between node $v_j$ and node $v_k$.
This property measures the average betweenness centrality of the nodes with degree $k$ \cite{mahadevan, orsini}.
\item Largest eigenvalue of an adjacency matrix $\bm{A}$, as denoted by $\lambda_1$.
\end{enumerate}
We regard properties (1)--(7) as local structural properties and properties (8)--(12) as global structural properties.
For the properties involving shortest paths (i.e., $\bar{l}$, $\{P(l)\}_l$, $l_{\text{max}}$, and $\{\bar{b}(k)\}_k$), we calculate those of the largest connected component of a given graph.
To reduce the simulation time, we use the parallel algorithms presented in Ref. \cite{bader2006} to calculate $\bar{l}, \{P(l)\}_l, l_{\text{max}},$ and $\{\bar{b}(k)\}_k$ of a given graph.
Note that the use of these parallel algorithms does not affect the performance of each method.

\subsection{Accuracy measure}
To measure the accuracy of the structural properties of a generated graph, we calculate the normalized $L^1$ distance for each of the 12 structural properties between the original and generated graphs, as in Ref. \cite{gjoka_2_5_k}.
For each structural property, we denote the vector representing the property of the original graph as $\bm{x}$ and the vector representing that of the generated graph as $\bm{\tilde{x}}$.
We define the normalized $L^1$ distance between $\bm{x}$ and $\bm{\tilde{x}}$ as $\sum_i |\tilde{x}_i - x_i|/\sum_i x_i$.
For example, the $L^1$ distance between the degree distribution of the original graph, $\{P(k)\}_k$, and that of the generated graph, as denoted by $\{\tilde{P}(k)\}_k$, is given by $\sum_{k} |\tilde{P}(k) - P(k)|/\sum_k P(k)$.
For the scalar properties (i.e., $n$, $\bar{k}$, $\bar{c}$, $l_{\text{max}}$, and $\lambda_1$), the $L^1$ distance is equivalent to the relative error.
For example, the $L^1$ distance between the number of nodes in the original graph, $n$, and that in the generated graph, as denoted by $\tilde{n}$, is given by $|\tilde{n} - n|/n$.

\subsection{Methods to be compared}
We compare our method with two existing methods.
\begin{itemize}
\item {\bf Subgraph sampling} \cite{ahn, mislove, kwak, wilson, ahmed, leskovec}. 
One constructs a subgraph induced from a set of edges obtained using a crawling method.
The crawling method is arbitrary. 
Therefore, we consider three well-used crawling methods in addition to a random walk (RW).
\begin{itemize}
\item Breadth-first search (BFS) \cite{mislove, wilson, kurant2011, catanese,  rozemberczki2020}. One selects a seed node and explores all of its neighbors. Then, one traverses the earliest explored node, and explores all of its neighbors that have not been traversed. One repeats this procedure.
\item Snowball sampling \cite{goodman1961, lee2006, ahn, illenberger2012, rozemberczki2020}. 
All the neighbors are not explored, unlike the BFS procedure, and at most $k$ neighbors are chosen randomly at every iteration.
\item Forest fire sampling (FF) \cite{leskovec, de2010, ahmed, rozemberczki2020}.
FF is a stochastic version of snowball sampling. 
At every iteration, one explores a random proportion of neighbors. The proportion is sampled from a geometric distribution with the mean $p_f/(1-p_f)$, where $p_f$ is a parameter. 
Note that this process can finish before a target fraction of nodes is sampled.
In this case, we uniformly randomly select a node from the sampled nodes and revive the process from the node, as in Ref. \cite{kurant2011}.
\end{itemize}
\item {\bf Gjoka et al.'s method} \cite{gjoka_2_5_k}. 
This method generates a graph that preserves the estimates of local structural properties.
This method does not use any structure of the subgraph sampled by a random walk.
Unfortunately, we found that it is difficult to reproduce the original method based on their paper and the source code \cite{gjoka_2_5_k}.
We describe how to implement the reproducible version of Gjoka et al.'s method in Appendix \ref{appendix:B}.
\end{itemize}

We apply each method in a single run as follows.
We first uniformly and randomly select a seed node from a set of nodes.
Then, we start BFS, snowball sampling, FF, and RW from the same seed.
We continue each sampling procedure until the percentage of queried nodes reaches a given value.
For subgraph sampling using RW, Gjoka et al's method, and the proposed method, we perform these methods for the same RW to achieve a fair comparison. 

\subsection{Parameters}
In snowball sampling, we set $k=50$, as in Ref. \cite{rozemberczki2020}.
In FF, we set $p_f = 0.7$, as in Ref. \cite{ahmed}.
In the proposed and Gjoka et al.'s methods, we set the coefficient of the number of rewiring attempts as $R_{\text{C}} = 500$, based on Ref. \cite{orsini}.

\begin{figure*}[t]
      \begin{minipage}{0.32\hsize}
        \begin{center}
          \includegraphics[scale=\figscalea]{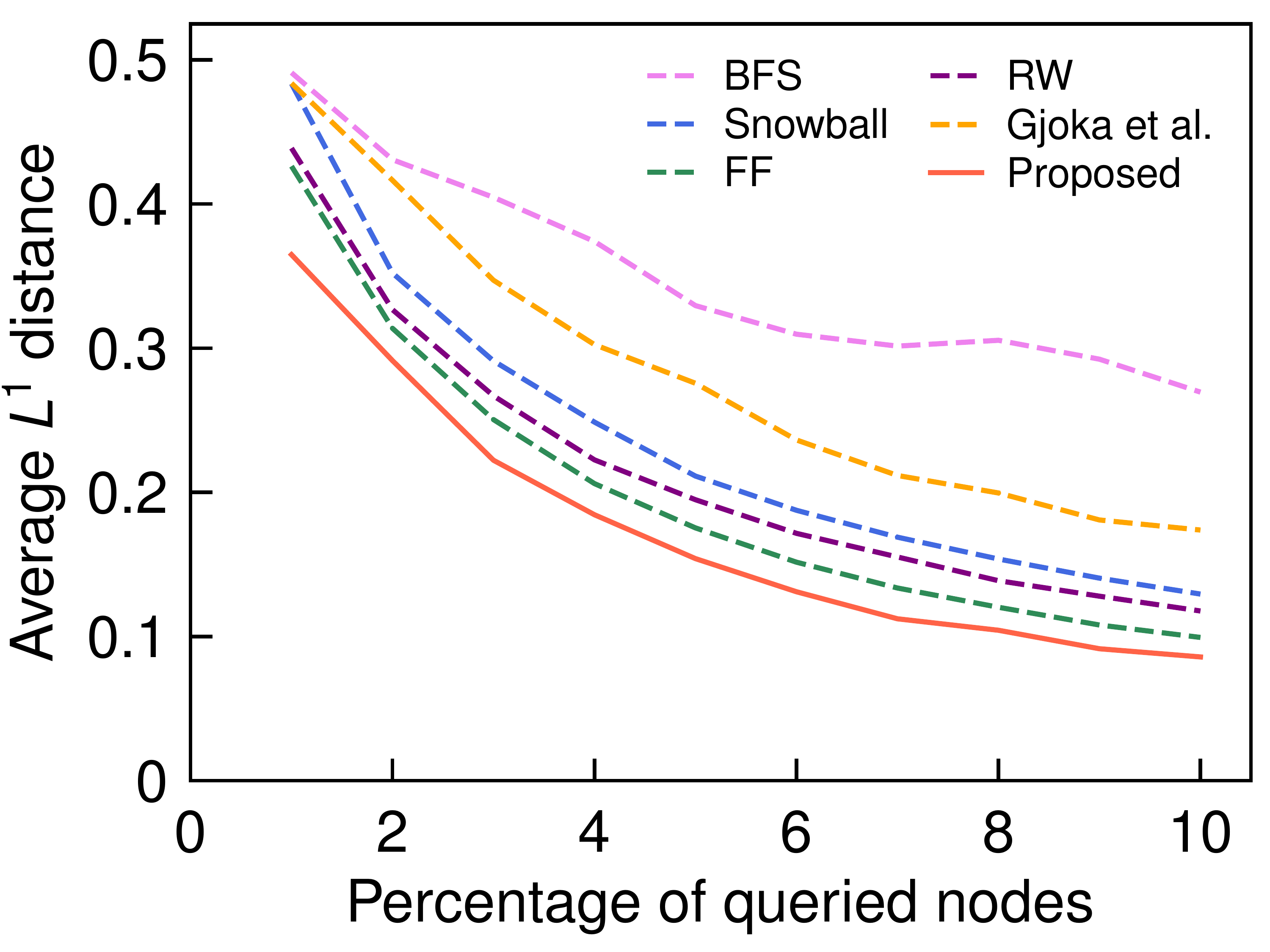}
          \\(a) Anybeat
        \end{center}
      \end{minipage}
      \hspace{2mm}
      \begin{minipage}{0.32\hsize}
        \begin{center}
          \includegraphics[scale=\figscalea]{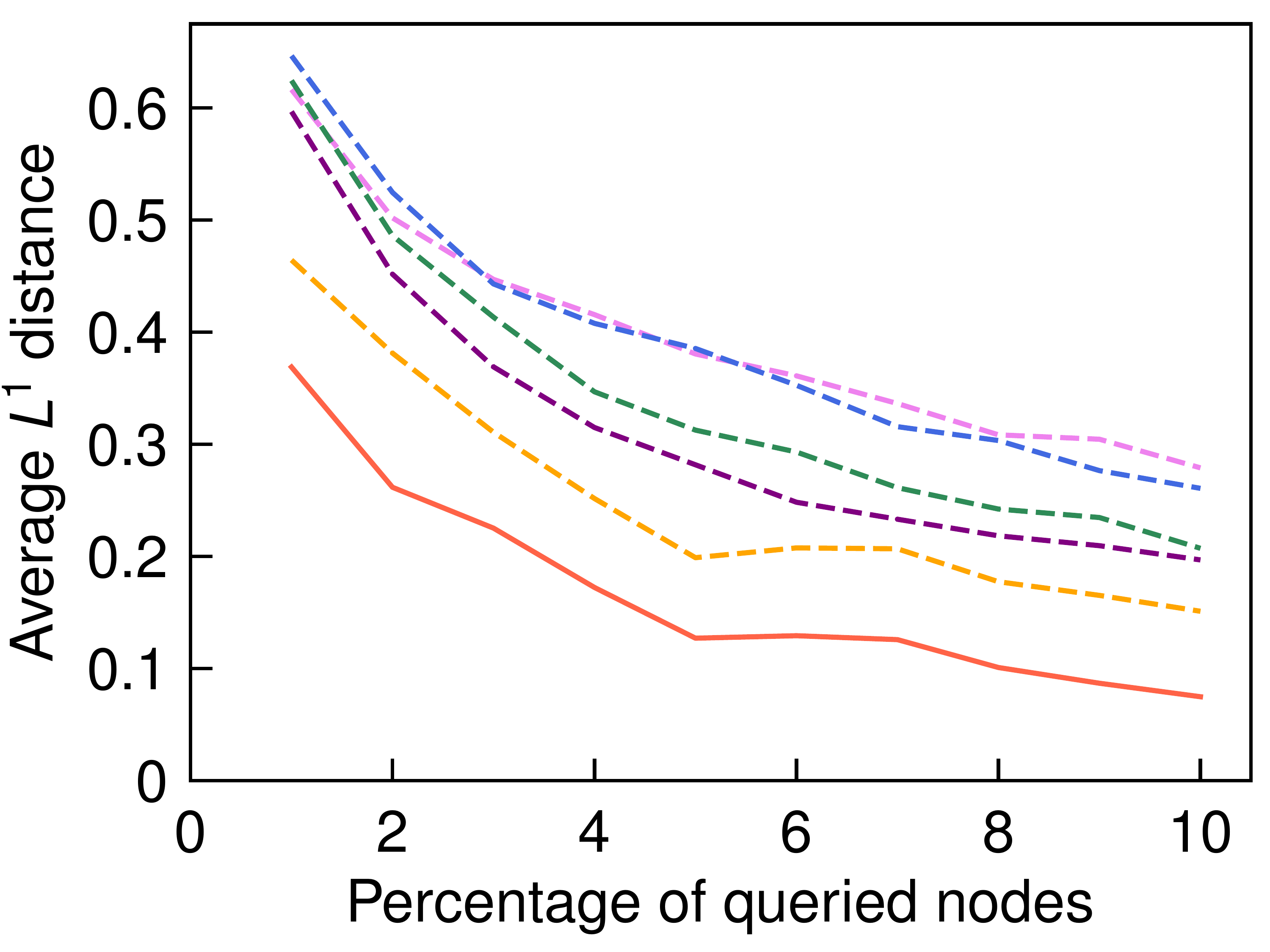}
          \\(b) Brightkite
        \end{center}
      \end{minipage}
      \hspace{2mm}
      \begin{minipage}{0.32\hsize}
        \begin{center}
          \includegraphics[scale=\figscalea]{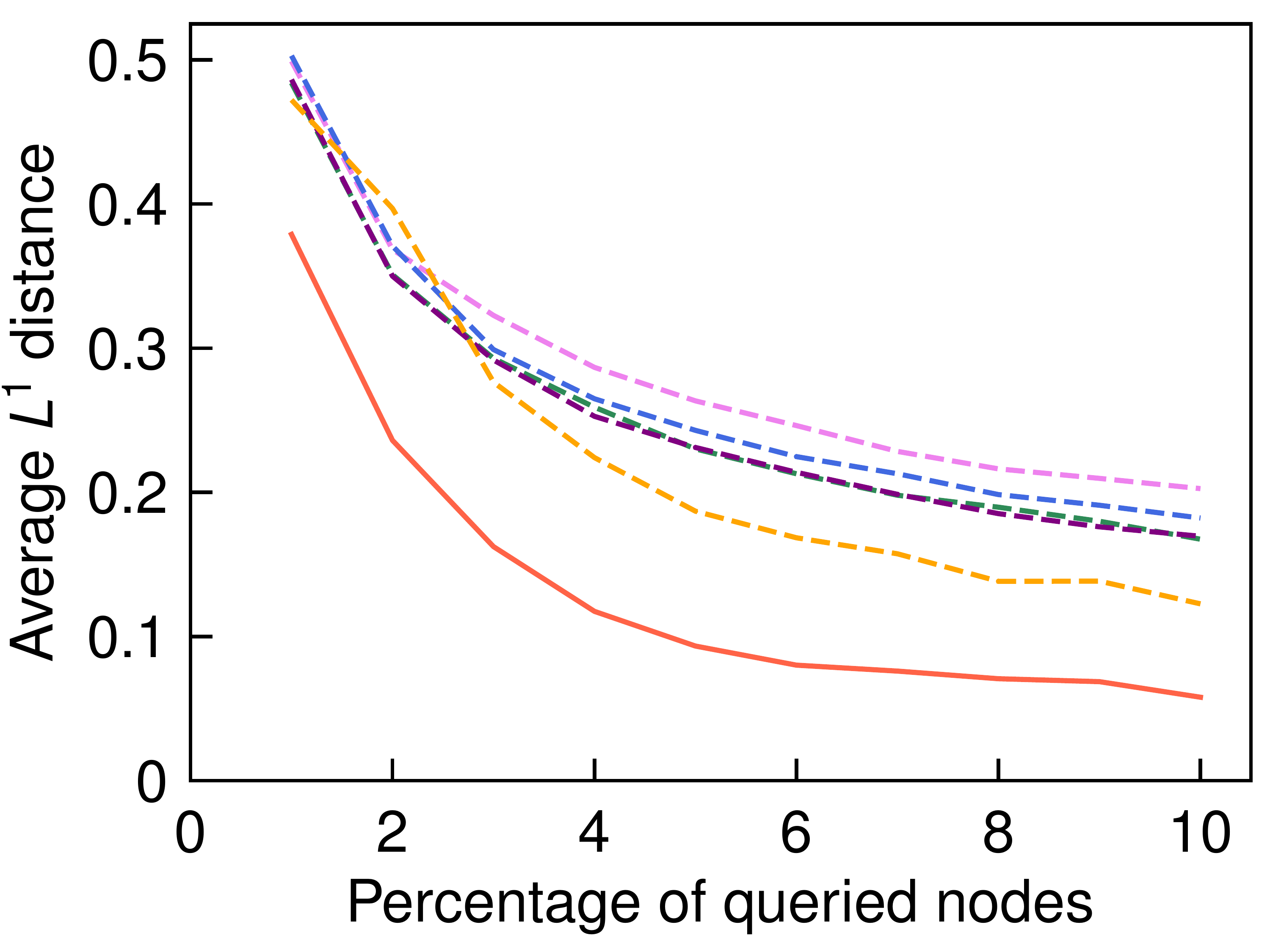}
          \\(c) Epinions
        \end{center}
      \end{minipage}
      \caption{Average $L^1$ distance over the 12 structural properties from different methods. We vary the percentage of queried nodes from 1\% to 10\% in increments of 1\%. All results are the average over 10 runs.}
      \label{fig:3}
      \vspace{-1mm}
\end{figure*}

\begin{table*}[t]
\caption{$L^1$ distance of each property from different methods using 10\% queried nodes. All results are the average over 10 runs. The lowest value is shown in bold.}
\label{table:2}
\vspace{-2mm}
\begin{center}
	\begin{tabular}{A | B | C C C C C C C C C C C C}\hline
	Dataset & Method & $n$ & $\bar{k}$ & $P(k)$ & $\bar{k}_{\text{nn}}(k)$ & $\bar{c}$ & $\bar{c}(k)$ & $P(s)$ & $\bar{l}$ & $P(l)$ & $l_{\text{max}}$ & $\bar{b}(k)$ & $\lambda_1$ \rule[0mm]{0mm}{3mm} \\ \hline
	\multirow{6}{*}{Slashdot} & BFS & 0.272 & 0.032 & 0.082 & 0.126 & 0.050 & 0.172 & 0.092 & 0.088 & 0.368 & 0.475 & 0.210 & 0.017 \\
	& Snowball & 0.248 & 0.043 & 0.074 & 0.102 & 0.057 & 0.152 & 0.092 & 0.086 & 0.356 & 0.392 & 0.108 & 0.013 \\
	& FF & 0.237 & 0.042 & 0.073 & 0.102 & 0.029 & 0.164 & 0.095 & 0.083 & 0.349 & 0.300 & 0.094 & 0.014 \\
	& RW & 0.242 & 0.042 & 0.072 & 0.102 & {\bf 0.023} & {\bf 0.150} & 0.099 & 0.084 & 0.352 & 0.225 & 0.100 & {\bf 0.011} \\
	& Gjoka et al. & {\bf 0.026} & {\bf 0.024} & 0.057 & 0.100 & 0.097 & 0.708 & 0.353 & {\bf 0.018} & {\bf 0.091} & {\bf 0.033} & 0.258 & 0.016 \\
	& Proposed & 0.029 & 0.027 & {\bf 0.056} & {\bf 0.042} & 0.097 & 0.205 & {\bf 0.034} & 0.025 & 0.101 & 0.058 & {\bf 0.068} & {\bf 0.011} \\ \hline
	\multirow{6}{*}{Gowalla} & BFS & 0.432 & 0.100 & 0.356 & 0.324 & 0.234 & 0.153 & 0.098 & 0.203 & 0.851 & 0.556 & 0.337 & 0.009\rule[0mm]{0mm}{2.5mm} \\
	& Snowball & 0.442 & 0.038 & 0.323 & 0.191 & 0.154 & 0.095 & {\bf 0.080} & 0.168 & 0.627 & 0.413 & 0.244 & 0.007 \\
	& FF & 0.408 & 0.032 & 0.280 & 0.149 & 0.102 & 0.072 & 0.087 & 0.140 & 0.511 & 0.256 & 0.176 & 0.008 \\
	& RW & 0.395 & 0.040 & 0.273 & 0.133 & 0.072 & {\bf 0.064} & 0.099 & 0.137 & 0.500 & 0.244 & {\bf 0.154} & 0.006 \\
	& Gjoka et al. & 0.034 & 0.017 & {\bf 0.040} & 0.102 & {\bf 0.029} & 0.350 & 0.539 & 0.038 & 0.160 & {\bf 0.106} & 0.767 & 0.469 \\
	& Proposed & {\bf 0.032} & {\bf 0.015} & 0.041 & {\bf 0.047} & 0.273 & 0.110 & 0.142 & {\bf 0.031} & {\bf 0.116} & {\bf 0.106} & 0.250 & {\bf 0.001} \\ \hline
	\multirow{6}{*}{Livemocha} & BFS & 0.062 & 0.405 & 0.500 & 0.295 & 0.322 & 0.343 & 0.040 & 0.037 & 0.204 & {\bf 0.000} & 0.614 & 0.111\rule[0mm]{0mm}{2.5mm} \\
	& Snowball & 0.037 & 0.286 & 0.352 & 0.209 & 0.273 & 0.262 & {\bf 0.033} & 0.018 & 0.106 & 0.017 & 0.301 & 0.057 \\
	& FF & {\bf 0.025} & 0.262 & 0.324 & 0.179 & 0.279 & {\bf 0.233} & 0.039 & 0.015 & 0.088 & {\bf 0.000} & 0.212 & 0.044 \\
	& RW & 0.027 & 0.271 & 0.335 & 0.181 & 0.287 & 0.244 & 0.041 & 0.015 & 0.091 & {\bf 0.000} & 0.228 & 0.045 \\
	& Gjoka et al. & 0.033 & 0.021 & {\bf 0.126} & 0.142 & {\bf 0.097} & 0.883 & 0.619 & 0.012 & 0.114 & 0.783 & 0.386 & {\bf 0.012} \\
	& Proposed & 0.048 & {\bf 0.020} & 0.132 & {\bf 0.092} & 0.129 & 0.388 & 0.150 & {\bf 0.006} & {\bf 0.032} & 0.050 & {\bf 0.118} & 0.020 \\ \hline
  	\end{tabular}
\end{center}
\end{table*}

\begin{table*}[t]
\caption{Average and standard deviation of the $L^1$ distance for the 12 properties from different methods using 10\% queried nodes. Each result is shown as the $\text{average} \pm \text{standard deviation}$. All results are the average over 10 runs. The lowest value is shown in bold.}
\label{table:3}
\vspace{-2mm}
\begin{center}
	\begin{tabular}{D | E E E E E E}\hline
	Dataset & BFS & Snowball & FF & RW & Gjoka et al. & Proposed \\ \hline
	Anybeat & 0.270 $\pm$ 0.144 & 0.129 $\pm$ 0.098 & 0.099 $\pm$ 0.071 & 0.118 $\pm$ 0.082 & 0.174 $\pm$ 0.143 & {\bf 0.086} $\pm$ {\bf 0.062}\rule[0mm]{0mm}{2.5mm} \\
	Brightkite & 0.279 $\pm$ 0.205 & 0.261 $\pm$ 0.189 & 0.207 $\pm$ 0.165 & 0.197 $\pm$ 0.159 & 0.151 $\pm$ 0.159 & {\bf 0.075} $\pm$ {\bf 0.061} \\
	Epinions & 0.203 $\pm$ 0.192 & 0.182 $\pm$ 0.173 & 0.167 $\pm$ 0.156 & 0.170 $\pm$ 0.164 & 0.123 $\pm$ 0.132 & {\bf 0.058} $\pm$ {\bf 0.055} \\
	Slashdot & 0.165 $\pm$ 0.140 & 0.144 $\pm$ 0.118 & 0.132 $\pm$ 0.105 & 0.125 $\pm$ 0.098 & 0.148 $\pm$ 0.198 & {\bf 0.063} $\pm$ {\bf 0.057} \\
	Gowalla & 0.305 $\pm$ 0.223 & 0.232 $\pm$ 0.180 & 0.185 $\pm$ 0.147 & 0.176 $\pm$ 0.145 & 0.221 $\pm$ 0.242 & {\bf 0.097} $\pm$ {\bf 0.089} \\
	Livemocha & 0.244 $\pm$ 0.199 & 0.162 $\pm$ 0.126 & 0.142 $\pm$ 0.113 & 0.147 $\pm$ 0.118 & 0.269 $\pm$ 0.320 & {\bf 0.099} $\pm$ {\bf 0.105} \\ \hline
  	\end{tabular}
\end{center}
\end{table*}

\section{Experimental Results}
\subsection{Accuracy of structural properties}
Figure \ref{fig:3} shows the average $L^1$ distance over the 12 structural properties from different methods for the Anybeat, Brightkite, and Epinions datasets, with the percentage of queried nodes varying from 1\% to 10\%.
We observe that the proposed method outperforms the compared methods in terms of the average $L^1$ distance for all the percentages of queried nodes.
Specifically, the proposed method improves the lowest average $L^1$ distance among the compared methods by 13.1\%, 50.5\%, and 52.8\% (i.e., from 0.099 to 0.086, from 0.151 to 0.075, and from 0.123 to 0.058, respectively) using 10\% queried nodes on the Anybeat, Brightkite, and Epinions graphs, respectively.

Table \ref{table:2} shows the $L^1$ distance for each structural property from different methods using 10\% queried nodes on the Slashdot, Gowalla, and Livemocha datasets. 
We first compare the proposed method with subgraph sampling using BFS, snowball, FF, and RW.
First, the proposed method typically improves the $L^1$ distance for $n$, $\bar{k}$, $\{P(k)\}_k$, and $\{\bar{k}_{\text{nn}}(k)\}_k$.
Second, the proposed method typically worsens the $L^1$ distance for $\bar{c}$ and $\{\bar{c}(k)\}_k$.
This is because the generated graph does not exactly preserves the estimate of the node clustering due to the rewiring process.
Third, in many cases, the proposed method improves the $L^1$ distance for global properties, i.e., $\bar{l}$, $\{P(l)\}_l$, $l_{\text{max}}$, $\{\bar{b}(k)\}_k$, and $\lambda_1$.

We then compare the proposed method with Gjoka et al.'s method.
First, the proposed method achieves comparable or sometimes better $L^1$ distances for $n$, $\bar{k}$, $\{P(k)\}_k$, and $\{\bar{k}_{\text{nn}}(k)\}_k$.
This stems from our design of the algorithms for constructing the target degree vector and the target joint degree matrix while minimizing their errors relative to the original estimates.
Second, the proposed method improves the $L^1$ distance for $\{\bar{c}(k)\}_k$.
This is because the proposed method is more likely to succeed in the rewiring of edges to approach the estimate of $\{\hat{\bar{c}}(k)\}_k$ because the edges in the sampled subgraph are excluded from the candidate edges to be rewired. 
Third, the proposed method improves the $L^1$ distance for $\{P(s)\}_s$ but worsens that for $\bar{c}$. 
Fourth, the proposed method often improves the $L^1$ distance for global properties, i.e., $\bar{l}$, $\{P(l)\}_l$, $l_{\text{max}}$, $\{\bar{b}(k)\}_k$, and $\lambda_1$.

Table \ref{table:3} shows the average and standard deviation of the $L^1$ distance for the 12 properties from different methods using 10\% queried nodes.
The proposed method achieves the lowest average and standard deviation for the six datasets.

\begin{figure*}[t]
	  \begin{minipage}{0.24\hsize}
        \begin{center}
          \includegraphics[height=\figheightvisori]{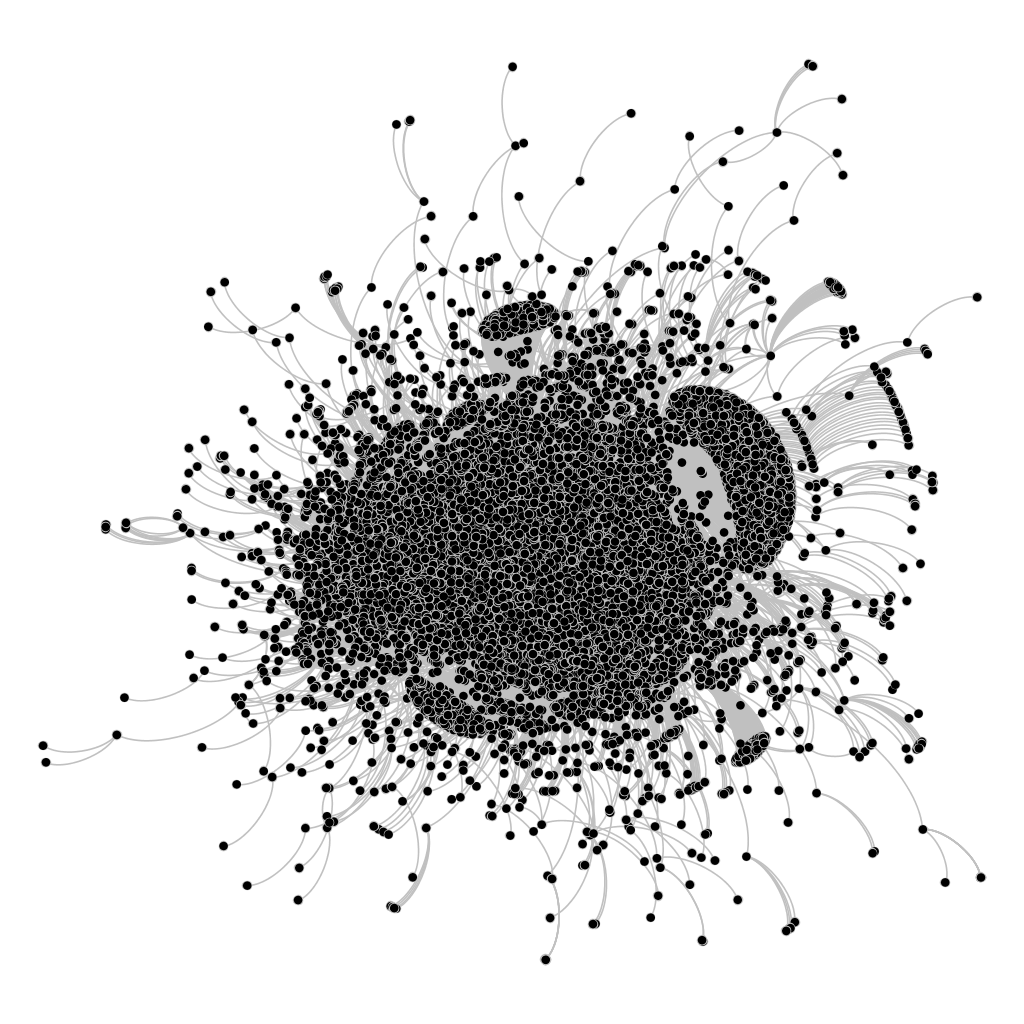}
          \\(a) Original graph
        \end{center}
      \end{minipage}
      \hspace{3mm}
      \begin{minipage}{0.22\hsize}
        \begin{center}
          \includegraphics[height=\figheightvis]{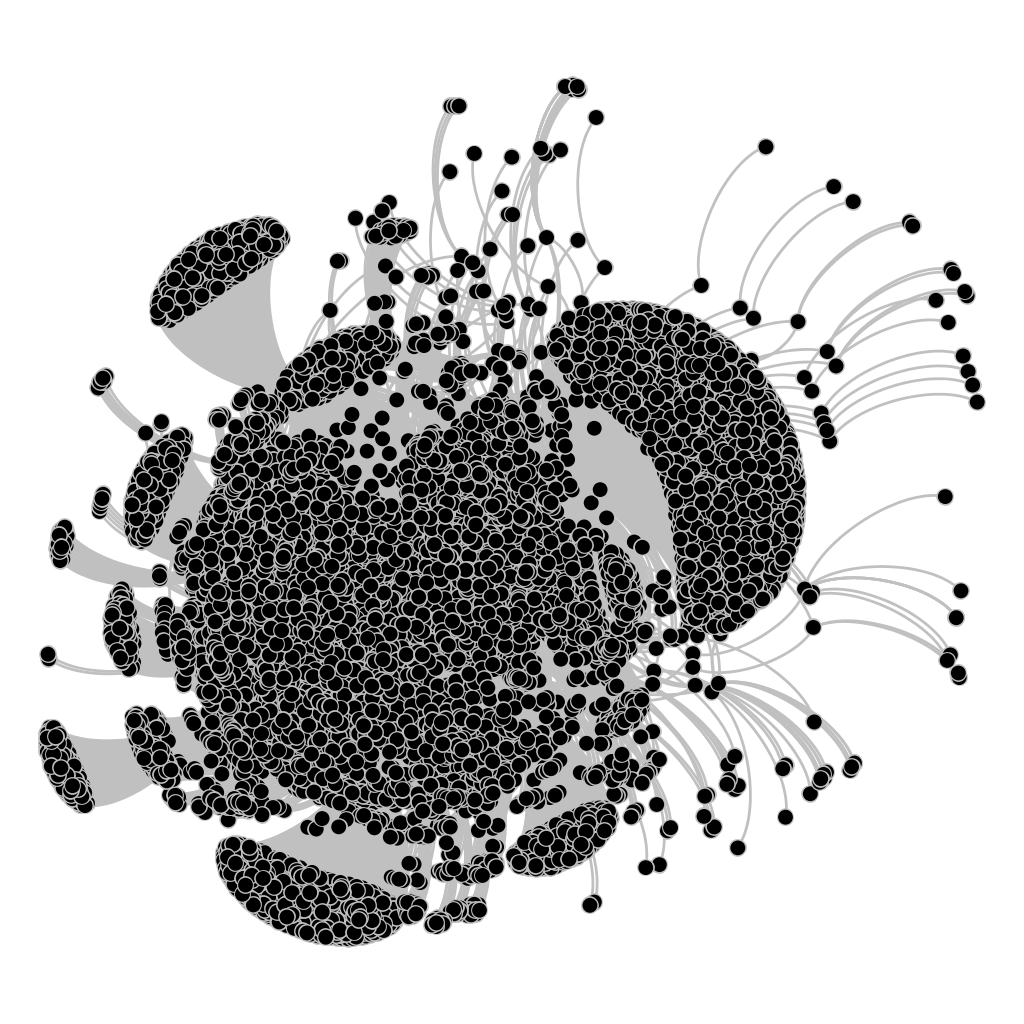}
          \\(b) BFS
          \\ \vspace{3mm}
          \includegraphics[height=\figheightvis]{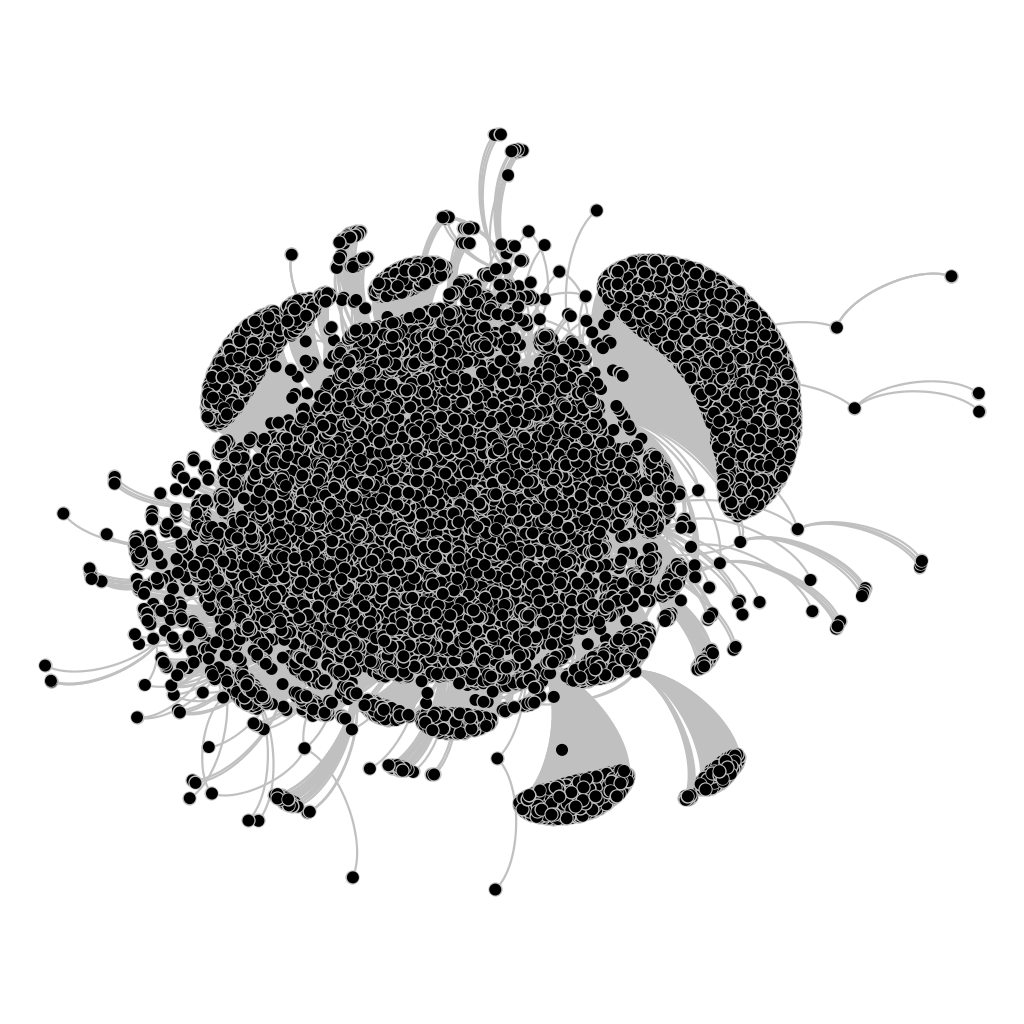}
          \\(e) RW
        \end{center}
      \end{minipage}
      \hspace{3mm}
      \begin{minipage}{0.22\hsize}
        \begin{center}
          \includegraphics[height=\figheightvis]{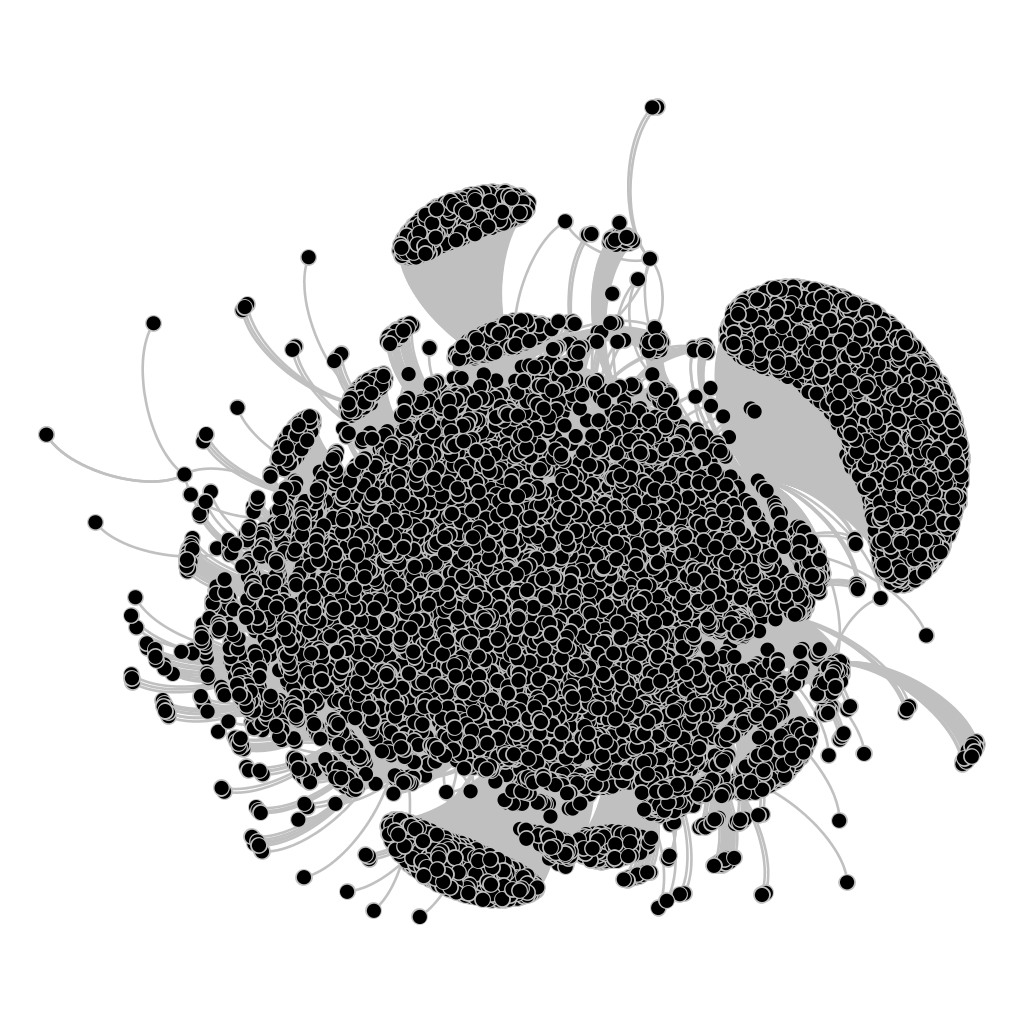}
          \\(c) Snowball
          \\ \vspace{3mm}
          \includegraphics[height=\figheightvis]{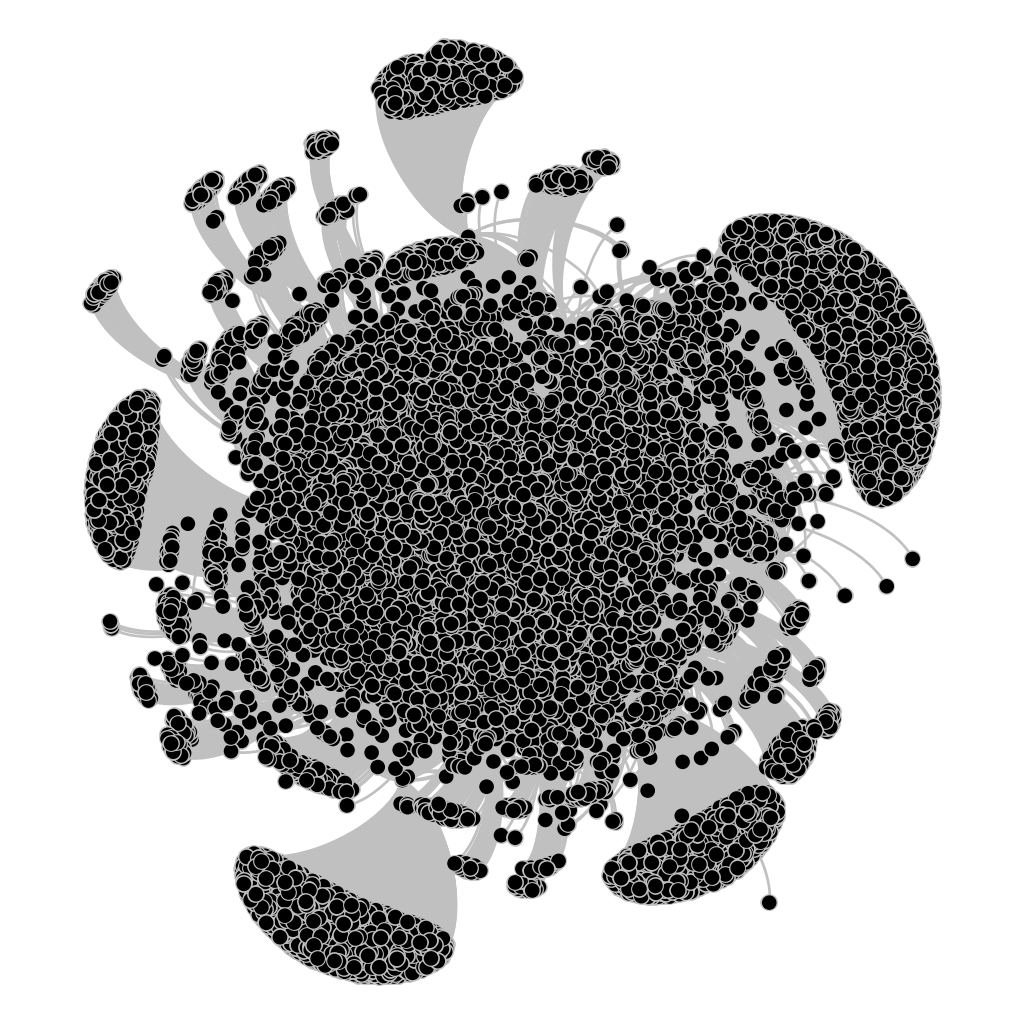}
          \\(f) Gjoka et al.
        \end{center}
      \end{minipage}
      \hspace{3mm}
      \begin{minipage}{0.22\hsize}
        \begin{center}
          \includegraphics[height=\figheightvis]{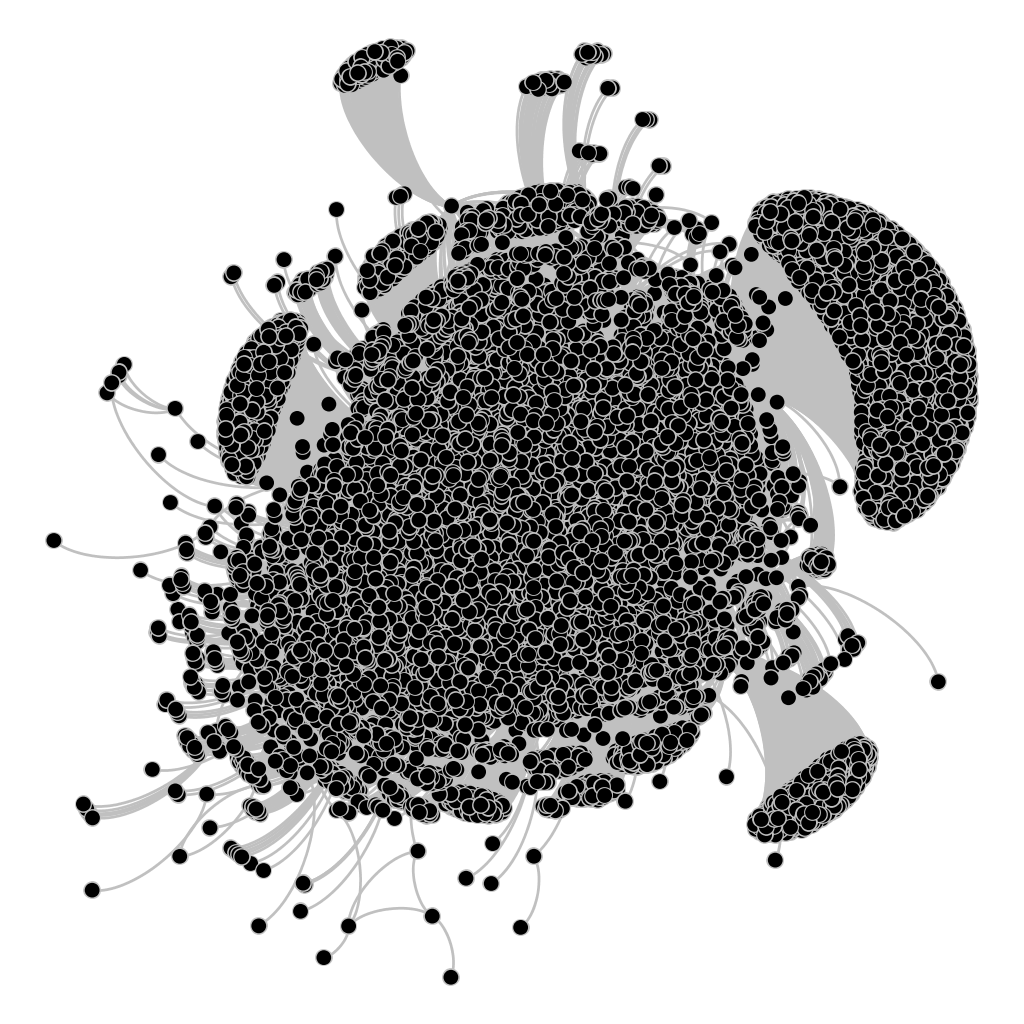}
          \\(d) FF
          \\ \vspace{3mm}
          \includegraphics[height=\figheightvis]{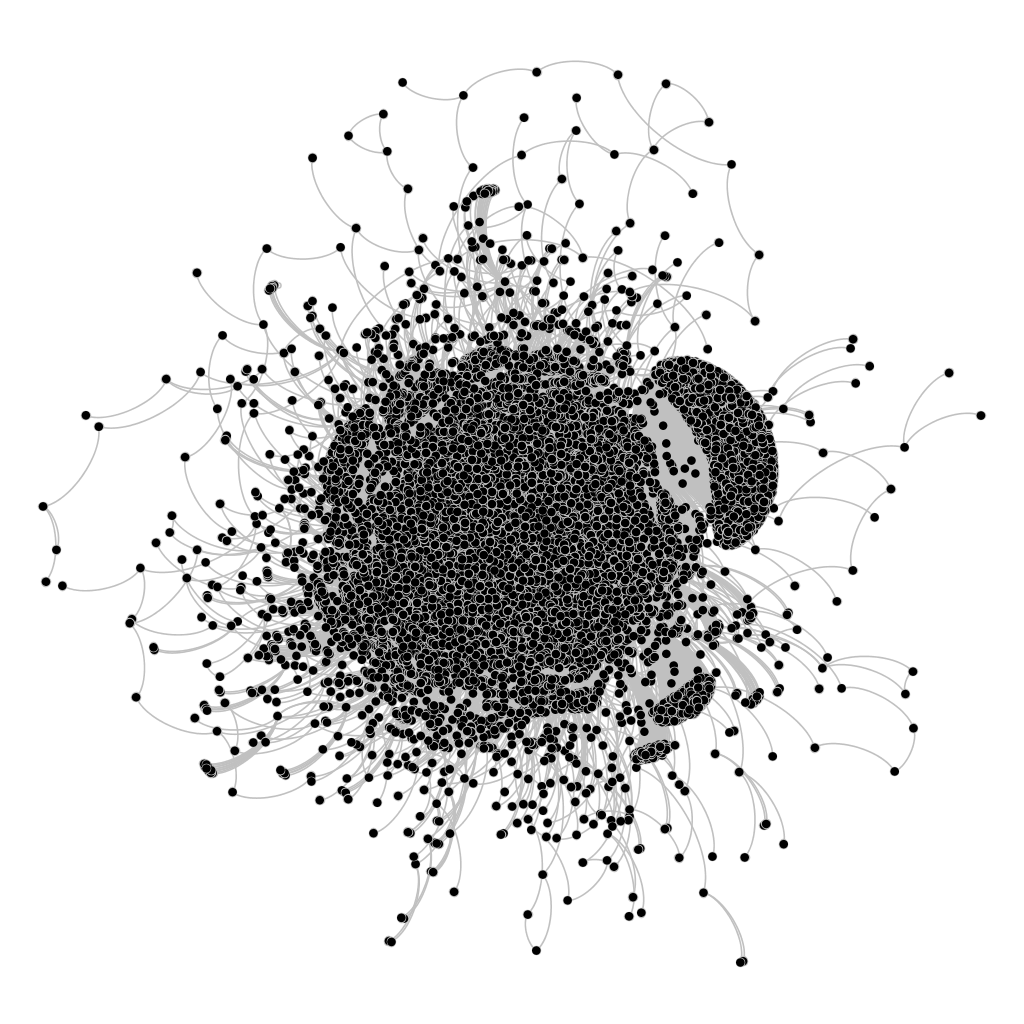}
          \\(g) Proposed
        \end{center}
      \end{minipage}
      \caption{Graph visualization for the Anybeat dataset. (a) Original graph. (b)--(g) Graphs generated by each method using 10\% queried nodes. The black circles represent nodes and the gray curves represents edges.}
      \label{fig:4}
\vspace{-1.5mm}
\end{figure*}

\begin{table*}[t]
\caption{Generation times (in seconds) of the different methods using 10\% queried nodes. For the proposed and Gjoka et al.'s methods, the total time and rewiring time are shown. All results are an average of 10 runs. The lowest value is shown in bold.}
\label{table:4}
\vspace{-1mm}
\begin{center}
	\begin{tabular}{D | I I I I I I I I}\hline
	\multirow{2}{*}{Dataset} & \multirow{2}{*}{BFS} & \multirow{2}{*}{Snowball} & \multirow{2}{*}{FF} & \multirow{2}{*}{RW} & \multicolumn{2}{c}{Gjoka et al.} & \multicolumn{2}{c}{Proposed} \\
	& & & & & Total & Rewiring & Total & Rewiring \\ \hline
	Anybeat & {\bf 0.008} & 0.014 & 0.014 & 0.018 & 547 & 546 & 60.6 & 56.4\rule[0mm]{0mm}{2.5mm} \\
	Brightkite & {\bf 0.120} & 0.121 & 0.142 & 0.148 & 703 & 694 & 192 & 163 \\
	Epinions & {\bf 0.235} & 0.248 & 0.243 & 0.259 & 2,914 & 2,878 & 280 & 168 \\
	Slashdot & {\bf 0.292} & 0.305 & 0.307 & 0.310 & 3,086 & 3,064 & 362 & 296 \\
	Gowalla & {\bf 0.954} & 1.08 & 1.19 & 1.30 & 15,907 & 15,735 & 4,255 & 3,679 \\
	Livemocha & {\bf 1.71} & 2.22 & 2.07 & 2.29 & 59,114 & 59,054 & 8,645 & 8,331 \\ \hline
  	\end{tabular}
\end{center}
\vspace{-2mm}
\end{table*}

\begin{table*}[t]
\caption{Performance of different methods using 1\% queried nodes for the YouTube graph. The $L^1$ distance for each property, the average (AVG) and standard deviation (SD) of the $L^1$ distance for 12 properties, and the generation time (in seconds) are shown. All results are an average of 5 runs. The lowest value is shown in bold.}
\label{table:5}
\vspace{-2mm}
\begin{center}
	\begin{tabular}{B | F F F F F F F F F F F F | G | H}\hline
	Method & $n$ & $\bar{k}$ & $P(k)$ & $\bar{k}_{\text{nn}}(k)$ & $\bar{c}$ & $\bar{c}(k)$ & $P(s)$ & $\bar{l}$ & $P(l)$ & $l_{\text{max}}$ & $\bar{b}(k)$ & $\lambda_1$ & AVG $\pm$ SD & Time (sec)\rule[0mm]{0mm}{3mm} \\ \hline
	BFS & 0.752 & 0.039 & 0.191 & 0.661 & 0.630 & 0.531 & 0.173 & 0.296 & 1.31 & 0.700 & 0.796 & 0.058 & 0.512 $\pm$ 0.363 & {\bf 0.724}\rule[0mm]{0mm}{2.5mm} \\
	Snowball & 0.749 & 0.060 & 0.180 & 0.608 & 0.620 & 0.593 & 0.131 & 0.263 & 1.11 & 0.625 & 0.792 & 0.074 & 0.484 $\pm$ 0.323 & 0.819 \\
	FF & 0.642 & 0.088 & 0.160 & 0.450 & 0.500 & 0.471 & 0.196 & 0.237 & 1.02 & 0.508 & 0.487 & 0.036 & 0.400 $\pm$ 0.264 & 1.30  \\
	RW & 0.637 & 0.051 & 0.166 & 0.514 & 0.536 & 0.532 & 0.190 & 0.233 & 1.00 & 0.417 & 0.491 & 0.036 & 0.400 $\pm$ 0.268 & 1.22 \\
	Gjoka et al. & {\bf 0.062} & {\bf 0.025} & {\bf 0.033} & 0.255 & 0.025 & 0.707 & 0.361 & 0.067 & 0.232 & 0.250 & 0.566 & 0.563 & 0.262 $\pm$ 0.236 & 77,334 \\
	Proposed & {\bf 0.062} & {\bf 0.025} & {\bf 0.033} & {\bf 0.196} & {\bf 0.022} & {\bf 0.409} & {\bf 0.106} & {\bf 0.042} & {\bf 0.191} & {\bf 0.142} & {\bf 0.412} & {\bf 0.014} & {\bf 0.138} $\pm$ {\bf 0.139} & 43,567 \\ \hline
  	\end{tabular}
\end{center}
\vspace{-2mm}
\end{table*}

\subsection{Graph visualization}
We compare the visual representations of graphs generated by different methods.
Figure \ref{fig:4}(a) shows the original graph and Figs. \ref{fig:4}(b)--(g) show the visual representations of the graphs generated by each method using 10\% queried nodes for the Anybeat dataset.
We use Gephi software \cite{bastian} to visualize each graph.
We make the following observations.
First, all subgraphs constructed using the BFS, snowball, FF, and RW methods capture the core structure consisting of high-degree nodes but not the peripheral structure consisting of low-degree nodes (see Figs. \ref{fig:4}(b)--\ref{fig:4}(e)).
This is because crawling methods typically collect samples biased toward high-degree nodes \cite{gjoka_practical,gjoka_walking}.
Second, Gjoka et al.'s method hardly reproduces the visual representation of the original graph (see Fig. \ref{fig:4}(f)) because their method does not use any structures of the sampled subgraph.
Third, the proposed method successfully reproduces the original structure in the visualization (see Fig. \ref{fig:4}(g)) because the generated graph preserves the structure of the sampled subgraph.
Fourth, the proposed method successfully reproduces not only the core structure  but also the peripheral structure, which subgraph sampling does not reproduce.

\subsection{Generation time}
We compare the generation times of different methods.
Table \ref{table:4} shows the generation times (in seconds) of the different methods using 10\% queried nodes for six datasets. 
For the proposed and Gjoka et al.'s methods, we show both the total generation time and the running time of the rewiring process.
Subgraph sampling is much faster because the construction time of the subgraph is linearly proportional to the number of edges in the subgraph.
The proposed and Gjoka et al.'s methods require much longer generation times than subgraph sampling, mainly due to the process of rewiring edges.
However, interestingly, the proposed method is several times faster than Gjoka et al.'s method for all six datasets, e.g., 9.0 times faster for the Anybeat dataset and 10.4 times faster for the Epinions dataset.
This is because the proposed method reduces the running time of the process of rewiring edges, which is a bottleneck in the generation time, because our rewiring procedure excludes the edges in the sampled subgraph from the candidate edges to be rewired.
For the proposed and Gjoka et al.'s methods, although the rewiring time is reduced upon decreasing the coefficient of the number of rewiring attempts $R_{\text{C}}$, we note that the reproducibility of the structural properties, including clustering coefficients, of the generated graphs is also reduced.

\subsection{Performance on the YouTube dataset}
Finally, we compare the performance of the different methods using 1\% queried nodes on the YouTube dataset.
Table \ref{table:5} shows the $L^1$ distance for each property, the average and standard deviation of the $L^1$ distance for the 12 properties, and the generation time for each method.
First, the proposed method achieves the lowest $L^1$ distance for most of the 12 properties.
Second, the proposed method improves the average and standard deviation of the $L^1$ distance over the 12 properties by 47.3\% (from 0.262 to 0.138) and 41.1\% (from 0.236 to 0.139), respectively, compared with the lowest value among the existing methods.
Third, the generation time of the proposed method is reduced by 43.7\% compared to that of Gjoka et al.'s method.
As expected, the main factor in this increased speed is the reduction in the rewiring time; the proposed method requires 37,990 seconds for rewiring, whereas Gjoka et al.'s method requires 77,271 seconds.
Although subgraph sampling is considerably faster, the reproducibility of the structural properties of the subgraphs is much worse than that of graphs generated by the proposed method.

\section{Conclusion}
In this study, we proposed a method for restoring the original social graph from the small sample obtained by a random walk. 
The proposed method generates a graph that preserves estimates of local structural properties and the structure of the subgraph sampled by a random walk. 
We compared the proposed method with subgraph sampling and Gjoka et al.'s method in terms of the accuracy of 12 structural properties, the visual representation, and the generation time for generated graphs. 
We showed that the proposed method generates graphs that more accurately reproduce the structural properties on average and the visual representation of the original graph than the compared methods. 
Furthermore, the generation time of the proposed method is several times faster than that of Gjoka et al.'s method. 
If most of the graph data could be sampled (e.g., if 50\% or more of the nodes could be queried), subgraph sampling is more effective than the proposed method because the subgraph should be almost structurally equivalent to the original graph and its construction time is fast. 
However, it is often difficult to collect a large sample of social graphs in practical scenarios. 
For example, the percentage of queried nodes was less than $1\%$ in a case study of crawling the Facebook graph \cite{gjoka_practical, gjoka_walking}.
Based on these results, we suggest investing in methods to complement the nodes and edges in the subgraph sampled by a random walk to realize exhaustive analyses of social graphs with limited data access.

There are several future directions for this research.
The first is to study a method with theoretical guarantees for restoring the social graph.
The proposed method enables us to estimate various structural properties with good accuracy on average but has one limitation; i.e., there is no guarantee of error bounds in the structural properties for the generated graphs. 
This is mainly because the $dK$-series \cite{gjoka_2_5_k, mahadevan, orsini}, which is the family of generative models underlying the proposed method, does not guarantee error bounds in the structural properties of the generated graphs.
The second is to study a scalable restoration method to deal with large-scale social graphs.
The proposed method suffers from a considerably high computational overhead compared to subgraph sampling, although the proposed method is several times faster than Gjoka et al.'s method \cite{gjoka_2_5_k}.
This is mainly due to the process of rewiring edges in a generated graph and existing studies on the $dK$-series \cite{gjoka_2_5_k, mahadevan, orsini} also faced high computational costs due to the rewiring process.
Studying a restoration method based on scalable graph generative models that accurately reproduce the structural properties of a given graph could improve this limitation.
Finally, it would be interesting to use or extend the dissimilarity \cite{schieber2017} of a given graph to investigate how well the proposed method restores the original social graph.

\section*{Acknowledgment}
Kazuki Nakajima was supported by JSPS KAKENHI Grant Number JP21J10415.
Kazuyuki Shudo was supported by JSPS KAKENHI Grant Number JP21H04872.

\appendices

\section{Unbiasedness of the Joint Degree Distribution Estimator} \label{appendix:A}

In this section, we prove that the estimator $\hat{P}(k, k')$ proposed in \cite{gjoka_2_5_k} is an unbiased estimator of the joint degree distribution $P(k, k')$.

\begin{lemma}
$\hat{P}(k, k')$ is an asymptotically unbiased estimator of $P(k, k')$.
\end{lemma}

\begin{IEEEproof}
We show that both $\hat{P}_{\text{IE}}(k, k')$ and $\hat{P}_{\text{TE}}(k, k')$ are asymptotically unbiased estimators of $P(k, k')$.
First, we calculate the expectation of $\Phi(k, k')$ with respect to the stationary distribution of a simple random walk.
\begin{align*}
&\mathbb{E}[\Phi(k, k')] \\
= &\mathbb{E}\left[\frac{1}{k k'} 1_{\{d_{x_{i'}}=k \land d_{x_{j'}}=k'\}} A_{x_{i'}, x_{j'}} \right] \notag \\
= &\sum_{i=1}^n \sum_{j=1}^n \frac{d_i}{2m} \frac{d_j}{2m} \\
&\ \ \ \ \ \times \mathbb{E}\left[\frac{1}{k k'} 1_{\{d_{x_{i'}}=k \land d_{x_{j'}}=k'\}} A_{x_{i'}, x_{j'}} \middle | x_{i'}=i, x_{j'}=j \right] \notag \\
= &\frac{1}{4m^2} \sum_{i=1,\ d_i=k}^n \sum_{j=1,\ d_j=k}^n A_{i, j} \notag \\
= &\frac{1}{2m} P(k, k').
\end{align*}
The first equation holds because of the linearity of expectation.
The second equation holds for the following reasons: (i) the stationary distribution of a simple random walk in a state space of a set of nodes is given by $\pi_{\mathcal{V}} = (d_i/2m)_{i=1}^n$ \cite{levin}; (ii) we have the law of total expectation; and (iii) a node pair $v_{x_i'}$ and $v_{x_j'}$ is regarded as being independently sampled if those ordinal numbers in the sample sequence (i.e., $x_{i'}$ and $x_{j'}$) are sufficiently far apart \cite{katzir_node, katzir_nodecc}.
Then, we conclude that $\hat{P}_{\text{IE}}(k, k')$ is an asymptotically unbiased estimator of $P(k, k')$ because it holds that
\begin{align*}
\mathbb{E}[\hat{n}] \mathbb{E}[\hat{d}_{\text{avg}}] \mathbb{E}[\Phi(k, k')] &= n d_{\text{avg}} \frac{1}{2m} P(k, k') \\
& = P(k, k').
\end{align*}
The second equation holds because of the handshaking lemma.

Next, we calculate the expectation of $\hat{P}_{\text{TE}}(k, k')$:
\begin{align*}
&\mathbb{E}[\hat{P}_{\text{TE}}(k, k')] \\
= &\sum_{(v_i, v_j) \in \mathcal{E}} \frac{1}{2m} \left(1_{\{d_i=k \land d_j=k'\}} + 1_{\{d_i=k' \land d_j=k\}}\right) \notag \\
= &P(k, k').
\end{align*}
The first equation holds for the following reasons; (i) we have the linearity of expectation; (ii) the stationary distribution of a simple random walk in a state space of a set of edges is given by $\pi_{\mathcal{E}} = (1/2m)_{(v_i, v_j) \in \mathcal{E}}$ \cite{levin}; and (iii) we have the law of total expectation.
Therefore, we obtain the desired result.
\end{IEEEproof}

\section{Implementation of Gjoka et al.'s Method} \label{appendix:B}
In this section, we describe the implementation of Gjoka et al.'s method \cite{gjoka_2_5_k}.
The underlying idea of Gjoka et al.'s method is to generate a graph that preserves the estimates of local structural properties obtained by re-weighted random walk.
However, we found that it is difficult to reproduce the original method from their paper \cite{gjoka_2_5_k} and the source code. 
Therefore, we implemented a reproducible version of this method by utilizing the algorithms of the proposed method as follows.

First, we obtain the estimates of local structural properties (i.e., the number of nodes $\hat{n}$, average degree $\hat{\bar{k}}$, degree distribution $\{\hat{P}(k)\}_k$, joint degree distribution $\{\hat{P}(k, k')\}_{k, k'}$, and degree-dependent clustering coefficient $\{\hat{\bar{c}}(k)\}_k$) using re-weighted random walk (see Section III. E). 
Second, we construct the target degree vector.
To this end, we perform the initialization step described in Section IV. B and then perform the adjustment step described in Section IV. B.
We do not perform the modification step for the target degree vector because Gjoka et al.'s method does not use any structural information of the subgraph sampled by a random walk.
Third, we construct the target joint degree matrix.
To this end, we perform the initialization step described in Section IV. C and then perform the adjustment step described in Section IV. C.
We do not perform the modification step for the target joint degree matrix for the same reason given in the construction of the target degree vector.
Fourth, we construct a graph that preserves the target degree vector and the target joint degree matrix from an empty graph by using the existing procedure \cite{mahadevan, stanton}.
Finally, we perform a process of rewiring edges so that the final graph approximately preserves the estimate of the degree-dependent clustering coefficient $\{\hat{\bar{c}}(k)\}_k$.
The rewiring procedure is the same as that described in Section IV. E, except that all the edges in the generated graph are candidates of edges to be rewired (i.e., $\tilde{\mathcal{E}}_{\text{rew}} = \tilde{\mathcal{E}}$).

\bibliographystyle{IEEEtran}
\bibliography{icde2022_tech_rep_nakajima.bbl}

\end{document}